\documentclass[%
 reprint,
superscriptaddress,
 amsmath,amssymb,
 aps,
prb,
]{revtex4-2}

\usepackage{graphicx}
\usepackage{dcolumn}
\usepackage{bm}
\usepackage{xcolor}
\usepackage[utf8]{inputenc}
\usepackage{amsmath}
\usepackage{amssymb}
\usepackage{graphicx}
\usepackage{gensymb}
\usepackage{graphics}
\usepackage[T1]{fontenc}
\usepackage{soul}
\usepackage[para,online,flushleft]{threeparttable}
\PassOptionsToPackage{hyphens}{url}\usepackage{hyperref}

\begin{document}

\preprint{APS/123-QED}

\title{NbSe$_2$'s charge density wave collapse in the (LaSe)$_{1.14}$(NbSe$_2$)$_2$ misfit layer compound}

\author{Ludovica Zullo}
\altaffiliation{These authors contributed equally to this work}
\email[]{\\ludovica.zullo@unitn.it}
\affiliation{Department of Physics, University of Trento, Via Sommarive 14, 38123 Povo, Italy}
\affiliation{Sorbonne Universit\'e, CNRS, Institut des Nanosciences de Paris, UMR7588, F-75252 Paris, France}

\author{Grégory Setnikar}
\altaffiliation{These authors contributed equally to this work}
\affiliation{CNRS, Universit\'e Grenoble Alpes, Institut N\'eel, 38042 Grenoble, France}

\author{Amit Pawbake}
\affiliation{CNRS, Universit\'e Grenoble Alpes, Institut N\'eel, 38042 Grenoble, France}
\author{Tristan Cren}
\affiliation{Sorbonne Universit\'e, CNRS, Institut des Nanosciences de Paris, UMR7588, F-75252 Paris, France}
\author{Christophe Brun}
\affiliation{Sorbonne Universit\'e, CNRS, Institut des Nanosciences de Paris, UMR7588, F-75252 Paris, France}
\author{Justine Cordiez}
\affiliation{Nantes Universit\'e, CNRS, Institut des Mat\'eriaux de Nantes Jean Rouxel, IMN, F-44000 Nantes, France}
\author{Shunsuke Sasaki}
\affiliation{Nantes Universit\'e, CNRS, Institut des Mat\'eriaux de Nantes Jean Rouxel, IMN, F-44000 Nantes, France}
\author{Laurent Cario}
\affiliation{Nantes Universit\'e, CNRS, Institut des Mat\'eriaux de Nantes Jean Rouxel, IMN, F-44000 Nantes, France}
\author{Giovanni Marini}
\affiliation{Department of Physics, University of Trento, Via Sommarive 14, 38123 Povo, Italy}
\author{Matteo Calandra}
\affiliation{Department of Physics, University of Trento, Via Sommarive 14, 38123 Povo, Italy}
\affiliation{Sorbonne Universit\'e, CNRS, Institut des Nanosciences de Paris, UMR7588, F-75252 Paris, France}
\author{Marie-Aude M\'easson}
\email[]{marie-aude.measson@neel.cnrs.fr}
\affiliation{CNRS, Universit\'e Grenoble Alpes, Institut N\'eel, 38042 Grenoble, France}

\date{\today}

\begin{abstract}
Misfit layer compounds, heterostructures composed by a regular alternating stacking of rocksalt monochalcogenides  bilayers and few-layer transition metal dichalchogenides, are an emergent platform to investigate highly doped transition metal dichalcogenides.
Among them, (LaSe)$_{1.14}$(NbSe$_2$)$_2$ displays Ising superconductivity, while the presence of a charge density wave (CDW) in the material is still under debate. Here, by using polarized Raman spectroscopy and first-principles calculations, we show that NbSe$_2$ undergoes a doping-driven collapse of the CDW ordering within the misfit, and no signature of the CDW is detected down to 8~K. 
We provide a complete experimental and theoretical description of the lattice dynamics of this misfit compound. We show that the vibrational properties are obtained from those of the two subunits, namely the LaSe unit and the NbSe$_2$ bilayer, in the presence of a suitable field-effect doping, and then highlight the 2D nature of the lattice dynamics of NbSe$_2$ within the (LaSe)$_{1.14}$(NbSe$_2$)$_2$ 3D structure.

\end{abstract}

\maketitle


\section{INTRODUCTION}

Transition metal dichalcogenides (TMDs) are a wide family of layered materials possessing fascinating physical phenomena \cite{MANZELI2017,FU2021,CHEN2022,JOSEPH2023}. 
Among these, bulk NbSe$_{2}$ displays competition between charge density wave (CDW) and superconducting order.  An incommensurate CDW transition at $33$ K occurs in bulk $2$H-NbSe$_{2}$  \cite{Wilson1975,Moncton1977a,Malliakas2013}. Superconductivity (SC) emerges below $7.2$ K and coexists with the CDW state\cite{Revolinsky1965}.
Recent experiments\cite{Xi2015,Ugeda2016} demonstrate that CDW survives in the two-dimensional (2D) limit for NbSe$_{2}$ bi and single layers. 

Achieving a complete control of CDW order in these systems could lead to a better understanding of the interplay between SC and CDW. To this aim, external parameters that can be tuned are doping, pressure, strain and sample thickness. However, each one of these  control knobs leads to different effects. For example, applying pressure to bulk NbSe$_2$ leads to a suppression of the CDW at $\approx 4.4$ GPa and 
an increase of the superconducting $T_c$   \cite{Moulding2020}, but no change in the ordering vector occurs. In the bulk, electron doping can be achieved via chemical intercalation \cite{Wang2020} paving the way to a tunability of the ordering vector via suitably chosen Fermi-nesting conditions. Exfoliation of 2D TMDs and ionic-liquid based field-effect transistors has led to the possibility of setting the doping electrochemically by tuning the voltage drop at the capacitor plates to generate an electrical-double layer in the proximity of the 2D dichalcogenide \cite{Novoselov2005a}. Experiments show that the CDW phase in bilayer NbSe$_{2}$ is weakened by electron doping \cite{Xi2016b}. This suggests that it could be possible to observe a CDW collapse at high voltages. 
Unfortunately, the amount of doping required to observe a collapse of the CDW phase exceeds the largest carrier chargings accessible via field effect gating (n$_{e}$ $\approx$ $3\times10^{14}$ e$^{-}$ cm$^{-2}$). Other approaches are thus needed to achieve higher doping.

Misfit layer compounds (MLCs) are an intriguing alternative for achieving nearly perfectly integrated 2D TMDs with massive doping \cite{ROUXEL1995,WIEGERS1996,Leriche2021,Zullo2023}. These heterostructures are formed by few-layer TMDs alternated with rocksalt units along the stacking direction.  The rocksalt units are electron donors and act as ultra efficient parallel plate capacitors\cite{Zullo2023} with a substantially boosted voltage drop at the rocksalt/TMD interface, much larger than the one achievable via conventional or electrical double-layer field-effect transistors. This leads to a massive electron charge transfer from the rocksalt to the TMDs.

We focus here on the misfit layer compound (LaSe)$_{1.14}$(NbSe$_2$)$_2$ that is composed of two subsystems, namely bilayers NbSe$_{2}$ (subsystem $1$) and LaSe rocksalt subunits (subsystem $2$) with different symmetries and periodicity \cite{Roesky1993}. The lattice parameter's mismatch along one of the in-plane direction of ratio $|\mathbf{a_{2}}|/|\mathbf{a_{1}}|=6 /3.437$ $(\approx7/4)$ makes  (LaSe)$_{1.14}$(NbSe$_2$)$_2$ an incommensurate compound. Quasiparticle interference measurements (QPIs) and angle-resolved photoemission spectroscopy (ARPES) show that each monolayer of NbSe$_2$ inside the (LaSe)$_{1.14}$(NbSe$_2$)$_2$ MLC is strongly electron-doped with a large Fermi level shift of $+0.3$ eV,  (corresponding to n$_{e}$ $\approx$ $6\times10^{14}$ e$^{-}$ cm$^{-2}$) \cite{Leriche2021}. Furthermore, scanning tunneling microscopy (STM) and magneto-transport measurements demonstrate that bulk (LaSe)$_{1.14}$(NbSe$_2$)$_2$ is superconducting at $5.7$ K with a critical field in the TMD plane that strongly violates the Pauli limit due to an efficient Ising protection, as in the monolayer case \cite{Samuely2021}.

Although superconductivity in (LaSe)$_{1.14}$(NbSe$_2$)$_2$ has been clearly demonstrated, the occurrence of CDW is still under debate. STM topography detected the presence of a short-range $2\times2$ modulation disappearing above $105$~K \cite{Leriche2021}. However, the $2\times2$ modulation observed in (LaSe)$_{1.14}$(NbSe$_2$)$_2$ by STM could be ascribed to a non-uniform doping on the cleaved surface. No bulk sensitive probes have demonstrated the presence or absence of a CDW in (LaSe)$_{1.14}$(NbSe$_2$)$_2$ up to now.

In this work, by performing Raman measurements and  first-principles electronic structure calculations, we demonstrate a CDW collapse in the NbSe$_2$ bilayers of (LaSe)$_{1.14}$(NbSe$_2$)$_2$ and ascribe it to the large electron transfer from the rocksalt to the TMD layers.
We assign the most intense Raman peaks to either of the MLC subunits (rocksalt or TMD) by comparing experimental Raman data and calculations. Finally, we show that, as it happens for the electronic structure\cite{Zullo2023}, the vibrational properties of MLC can be efficiently modeled by using a simple field-effect transistor scheme where each subunit can be seen as the gate of a parallel plate capacitor.
Our work sets a reference scheme for the interpretation of vibrational and structural properties of misfit layer compounds that can be extended to other compounds of the same family.

The paper is structured as follows: in section $2$ we describe the system and give the technical details of our experiment and first principles calculations.
In section $3$, we examine the theoretical rationale for the CDW stability of NbSe$_2$ inside the misfit. 
In section $4$ we discuss the collapse of CDW ordering in (LaSe)$_{1.14}$(NbSe$_2$)$_2$.
In section $5$ we present the Raman response of (LaSe)$_{1.14}$(NbSe$_2$)$_2$, and discuss the mode attribution in comparison with the theory, and, finally, in section $6$ we draw our conclusions.

\section{METHODS}

\subsection{Experiment}

\begin{figure}[]
\centering
\includegraphics[width=1\linewidth]{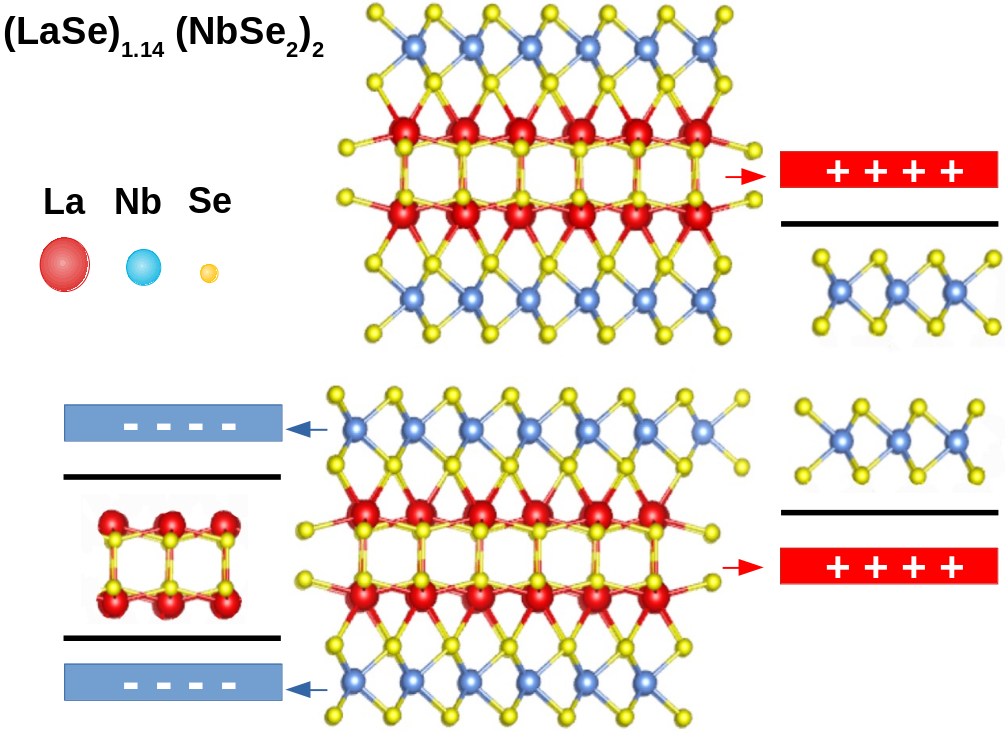}
\caption{
Crystal structure of (LaSe)$_{1.14}$(NbSe$_2$)$_2$ (center) composed of LaSe bilayers (red and yellow) and NbSe$_2$ bilayers (blue and yellow). Each LaSe bilayer donates $\approx 0.6$ electrons/Nb to each NbSe$_2$ layer. The NbSe$_2$ bilayer is modeled in a field effect transistor set up (right) in which each LaSe bilayer is replaced by a charged plate having a positive charge of $0.6$ electrons per Nb atom. The LaSe bilayer is modeled in a field effect transistor setup, replacing each NbSe$_2$ bilayer by a negatively charged plate of $0.6$ electrons per Nb.
To confine the atoms in the region in between the charged plates, positive potential barriers are added (black lines).
}

\label{fig1}
\end{figure}

Two single crystals of (LaSe)$_{1.14}$(NbSe$_2$)$_2$ were prepared by vapor transport using I$_2$, as detailed in Ref. \cite{Leriche2021}.  X-ray powder-diffraction experiments confirmed the so-called 1Q2H structure \cite{Leriche2021}, namely an alternated stacking of LaSe bilayers (NaCl structure) with TMDs bilayers of NbSe$_2$ (polytype 2H), as shown Fig. \ref{fig1}. 
Samples were freshly cleaved perpendicular to [001] axis just before performing Raman experiments in vacuum. 
Polarised Raman scattering has been performed in quasi-backscattering geometry with an incident laser line at 532~nm from a solid state laser. We used a closed-cycle $^{4}$He cryostat for the measurements from 8~K to 300~K. The scattered light was analysed by a Jobin Yvon T64000 triple subtractive grating spectrometer equipped with a cooled CCD detector. By keeping the optical set-up identical during the measurements, the Raman signal is quantitative. Particularly, the comparison of the intensity of the spectra between different temperatures is meaningful, and no normalisation has been applied. The contribution of the Bose factor has been removed from all spectra, with the temperatures corrected from the laser heating.

\subsection{Theory}

\subsubsection{Crystal structure} 

Given the incommensurability of the misfit layer compounds along one of the in-plane directions, a $(3+1)$D superspace group could be adopted to label the crystal structures \cite{WIEGERS1996}. However, the commensurate approximant of the (LaSe)$_1.14$(NbSe$_2$)$_2$ compound crystallizes in the $P1$ space group. The number of expected phonons is large, virtually infinite due to the incommensurability. From the symmetry analysis deriving from the $P1$ symmetry, all modes are both Raman (R) and Infrared (IR) active. However, many of the potentially Raman active  modes have very low intensity as (i) the symmetry of the subunits closely resemble the one of the isolated counterparts and (ii) the P$1$ group arises from the need of matching the rocksalt and TMD space groups. 

In order to gain a better understanding of the  Raman active phonon modes, we first consider the two sub-structures as separated, namely a rocksalt bilayer of LaSe and a bilayer of 2H-NbSe$_{2}$ (2L-NbSe$_2$). Along the c axis, one bilayer of NbSe$_2$ corresponds to one unit cell of 2H-NbSe$_2$.

In the presence of mirror symmetry with respect to the Nb plane (i.e. isolated NbSe$_2$ bilayer in the absence of an external electric field), the 2L-NbSe$_2$ sub-structure belongs to the space group P$\overline{3}$m$1$ (\#164, D$_{3d}^{3}$ point group). Each bilayer has 6 atoms per unit cell. The Wyckoff positions of the two Nb atoms are $2$c (with z$=3.13$\AA), while the four Se are in $2$d (with z$=1.47$\AA) and $2$d (with z$=4.84$\AA), respectively.

Bulk LaSe crystallizes in the Fm$\overline{3}$m (\#225) space group with two atoms per cell. However, we choose to label the atomic positions of the isolated LaSe bilayer by using the 
 $\rm Cmm2$ space group (\#35, C$_{2v}$ point group), which is suitable for the orthorhombic lattice of LaSe within the misfit. 
 The LaSe bilayer is an alternation of La and Se with a total of $8$ atoms per unit cell. The $4$ atoms composing the first layer have Wyckoff positions $2$a (with z$=-0.077$\AA) for Se and $2$b (with z$=-0.076$\AA) for La. The $4$ atoms composing the second layer have Wyckoff positions $2$a (with z$=0.076$\AA) for La and $2$b (with z$=0.077$\AA) for Se.
 
In our calculations, the in-plane lattice parameter of all the considered structures is fixed as the one of each sub-system in the bulk (LaSe)$_{1.14}$(NbSe$_2$)$_2$, namely, a$_{1}=3.437$ \AA \ and a$_{2}=6$ \AA \cite{Leriche2021}.

\subsubsection{Modeling of the bulk misfit}

Bulk (LaSe)$_{1.14}$(NbSe$_2$)$_2$ is a periodic arrangement of LaSe and NbSe$_{2}$ subunits along the stacking direction. The lattice parameter mismatch in one of the in-plane directions  makes the misfit cell incommensurate. It is possible to simulate an approximate commensurate cell \cite{Leriche2021}  by considering the ratio $|\mathbf{a_{2}}|/|\mathbf{a_{1}}|=6 /3.437$ $(\approx7/4)$, and thus  m =$7|\mathbf{a_{1}}| \approx 4|\mathbf{a_{2}}|$.  
This periodic approximant has been used to calculate the electronic structure  \cite{Leriche2021}, however it is still formed by too many atoms for the calculation of the vibrational properties.
In order to reduce the computational effort, we approximate the $7/4$ mismatch ratio by $8/4$, corresponding to a $2/1$ ratio. This is done by applying $14.6\%$ tensile strain to the rocksalt subunit, increasing the lattice parameter to a$_{2}=6.875$ \AA. The NbSe$_2$ in-plane parameter is, on the contrary, kept the same as in the misfit (a$_{1}=3.437$ \AA). 

Consequently, the two subunit cells in the $2\times 1$ periodic approximant of bulk (LaSe)$_{1.14}$(NbSe$_2$)$_2$ are listed below. The NbSe$_2$ sublattice has an orthorhombic cell with in-plane lattice vectors a$_{1}=3.437$ \AA \ and b$_{1}=6$ \AA, while the LaSe sublattice has an orthorhombic cell with in-plane lattice vectors a$_{2}=6.875$ \AA \ and b$_{2}=6$ \AA.

The resulting misfit crystal has an orthorhombic cell with lattice parameters a = $|\mathbf{b_{1}}|$ = $|\mathbf{b_{2}}|$ = $6$ \AA, \ b = $2|\mathbf{a_{1}}| \approx 1|\mathbf{a_{2}}|$=$6.875$ \AA \ and c= $18.25$ \AA. 
The structure has a P$1$ symmetry and includes $32$ atoms in the cell (atomic positions are reported in the Tab. 1 of the
Supplemental Material (SM)).
 
We calculate the vibrational properties of bulk (LaSe)$_{1.14}$(NbSe$_2$)$_2$ by means of  density functional perturbation theory (DFPT) as implemented in the
\textsc{quantum ESPRESSO}  (QE) code \cite{Giannozzi2020, Baroni2001} with ultrasoft pseudopotentials from pslibrary. 
The kinetic energy cutoff is set to 40 Ry and the Brillouin zone (BZ) integration is carried out over a 4$\times$4$\times$2 electron-momentum Monkhorst-Pack grid and by using a Gaussian smearing of $0.01$ Ry. The PBE \cite{Perdew1996} exchange and correlation functional is used in the calculations.

The atomic positions are fully optimised by means of the Broyden-Fletcher-Goldfarb-Shannon (BFGS) algorithm, with a convergence threshold of $10^{-4}$ Ry on the total energy difference between consecutive structural optimisation steps and of $10^{-3}$ Ry/Bohr on all forces components. 
During the relaxation procedure, we use the Van der Waals corrections Grimme-D3\cite{Grimme2011} to reproduce the interaction among adjacent NbSe$_{2}$ layers.  

We compute the dynamical matrix of bulk (LaSe)$_{1.14}$(NbSe$_2$)$_2$ at the $\Gamma$ point. The phonon density of states (PHDOS) is obtained by Fourier interpolation over a $10\times10\times1$ phonon-momentum grid and by using a Gaussian smearing of $3$ cm$^{-1}$. We note that in our calculations, the shearing mode along the axis with the lattice mismatch among the NbSe$_2$ and LaSe units goes slightly imaginary; nevertheless, this is an artifact caused by the tensile strain applied to the LaSe subunit.

\subsubsection{Modeling of the bulk misfit as a collection of field-effect transistors}

Inside the misfit, the LaSe subunit acts as a donor, losing $\approx$ 1.2 electrons and donating $\approx$ 0.6 electrons per Nb atom to each monolayer of the NbSe$_{2}$ bilayer subunit \cite{Leriche2021}. 
By means of the field effect transistor setup developed in Ref. \cite{Sohier2017,Brumme2015}, it is then possible to model the effect of the misfit structure onto the NbSe$_2$ bilayer by using a bilayer TMD sandwiched between two uniformly positive charged gates (see Fig. \ref{fig1}). Each charged gate replaces the RS subunit and has a positive charge per Nb corresponding to $0.6$ times the modulus of the electronic charge. This approach was efficiently carried out  
to estimate the misfit electronic structure in Ref. \cite{Zullo2023}.

The field-effect scheme can also be employed by considering an RS subunit sandwiched between two uniformly negative charged gates (see Fig. \ref{fig1}). In this case the goal is to determine the effect of the misfit structure onto the LaSe bilayer subunit so that the charged plates are now negatively charged.

The field-effect modeling is carried out by using  density functional theory (DFT) as implemented in the \textsc{quantum ESPRESSO} (QE) \cite{Giannozzi2020} package using the PBE exchange and correlation functional \cite{Perdew1996}.
We employ ultrasoft pseudopotentials from the Vanderbilt distribution for La and Nb, including semi-core states for Nb atoms \cite{Vanderbilt1990}, while for Se we use norm-conserving pseudopotentials with empty d-states in valence.
The kinetic energy cutoff for plane-wave basis set of NbSe$_{2}$ (LaSe) is set to 50 (48) Ry. The Brillouin zone (BZ) integration is performed with a Monkhorst-Pack grid of 21$\times$21$\times$1 (14$\times$14$\times$1) k-points and a Gaussian smearing of 0.01 (0.015) Ry. 

A Coulomb long range interaction cutoff is placed at z$_{cut}$ = $c/2$ with c being the unit-cell size in the direction perpendicular to the plane: c is set opportunely for each of the different systems to $20$ \AA \ for LaSe and $25$ \AA \ for NbSe$_{2}$. Each of the two subsystems is centred around z=$0$.
For 2L-NbSe$_{2}$ (LaSe) we use a double gate configuration, with two charged plates at z$_{bot}=-0.266c$ (z$_{bot}=-0.221c$) and z$_{top}=+0.266c$ (z$_{top}=+0.221c$) each with a charge of $\rho$=+0.6 ($\rho$=-0.6) times the modulus of the electronic charge, such that $\rho_{tot}=\rho_{2L}+\rho_{bot}+\rho_{top}=0$.
For each system a potential barriers V of height $2.5$ Ry is placed before the gates at z$_{V}$=z$_{bot}+0.1$ (z$_{V}$=z$_{top}-0.1$) in order to confine the atoms between the gate electrodes.

The Raman active phonon frequencies are calculated using density functional perturbation theory (DFPT) in the linear response regime \cite{Baroni2001}.
In order to fulfill the $7/4$ lattice mismatch ratio of the best periodic approximant,  the dynamical matrices are calculated on  uniform 7$\times$7$\times$1 and 4$\times$4$\times$1 phonon-momentum grids and then Fourier interpolated in the full Brillouin zone. For the DOS at zone center in Fig. \ref{fig4} panel b, we use only the phonon frequencies obtained from the dynamical matrices on a $4\times1\times 1$ phonon momentum grid. 

The individual phonon densities of states (PHDOS) in Fig. \ref{fig5} are obtained by Fourier interpolation over a $40\times40\times1$ and $70\times70\times1$ phonon-momentum grid for LaSe and NbSe$_{2}$ respectively, and by using a Gaussian smearing of $3$ cm$^{-1}$.

Vibrational properties of isolated neutral $2$L-NbSe$_2$ are calculated using DFPT in the linear response regime on  uniform 8$\times$8$\times$1 phonon-momentum grids. The Brillouin zone integration is performed with a Monkhorst-Pack grid of 30$\times$30$\times$1 k-points and a Methfessel-Paxton smearing of 0.005 Ry. 

\section{CDW stability in the misfit structure}

\begin{figure*}[t]
\centering
\includegraphics[width=0.95\linewidth]{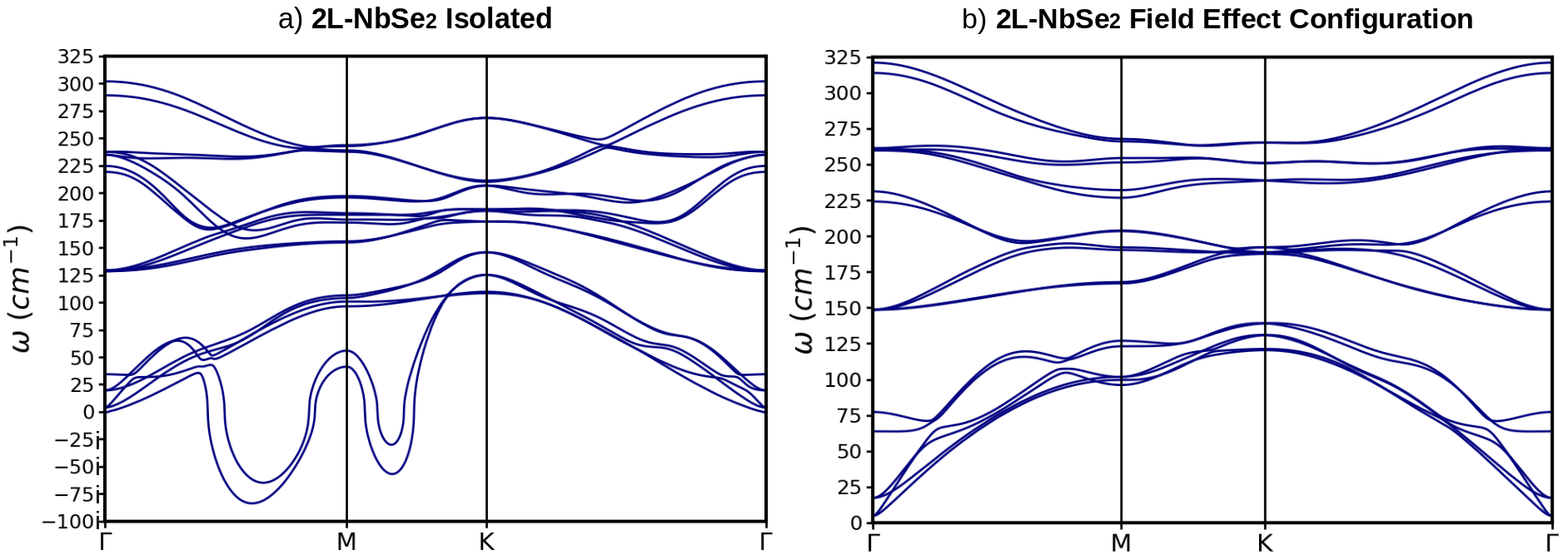}
\caption{Calculated phonon dispersion along the $\Gamma$-M-K-$\Gamma$ path  of a) an isolated $2$L-NbSe$_2$ and b) a field effect doped $2$L-NbSe$_2$. In panel b) the charge-density-wave instability is clearly removed by electron doping.}
\label{fig2}
\end{figure*}

In Fig. \ref{fig2} we calculate the harmonic phonon dispersion of an isolated neutral NbSe$_2$ bilayer (panel a) and of a NbSe$_2$ bilayer in field-effect configuration with a charging corresponding to $0.6$ electrons per niobium atom (panel b). The charge density wave instability occurring in the isolated NbSe$_2$ bilayer is showcased by the presence of an imaginary phonon band with the most imaginary value at ${\bf q}\approx 2/3 {\bf \Gamma M}$. Anharmonic effects do not qualitatively alter this behaviour, since the instability is reduced, but its wave vector is preserved \cite{Bianco2020}. In the FET charged NbSe$_2$ bilayer (panel b) the CDW instability is completely removed for charge transfers similar to those in the misfit. We thus expect that the CDW should collapse once the NbSe$_2$ subunit is inserted in the misfit. We will see that this  prediction is confirmed by Raman data.

We believe that our FET simulation can accurately recreate CDW behaviour as a function of misfit doping level. Indeed, in the first place, it has been demonstrated that the misfit generally behaves as a periodic arrangement of tunable field effect transistors \cite{Zullo2023}. In addition, in the specific case of (LaSe)$_{1.14}$(NbSe$_2$)$_2$, the electronic band structure of the misfit can be assimilated as that of a rigidly doped NbSe${_2}$ single layer \cite{Leriche2021}. Second, because the CDW in NbSe${_2}$ originates from the in-plane modes, FET modeling is appropriate for characterising its physics.

Finally, we conclude by noting that, if the charged plates mimicking the charge transfer by the LaSe subunits are removed and the FET charging is replaced by a uniform background doping, the results are completely different as they show an instability at the $M$ point, in qualitative disagreement with experiments (see for example supplemental materials of Ref. \cite{Leriche2021}). The reason is that in the misfit, as in a field-effect transistor, the charge transfer to the NbSe$_2$ bilayer is not uniformly distributed along the $c-$axis. For this reason, the uniform background doping approximation is inappropriate.

\section{Charge-transfer driven charge density wave collapse.}

\begin{figure}[]
\centering
\includegraphics[width=1\linewidth]{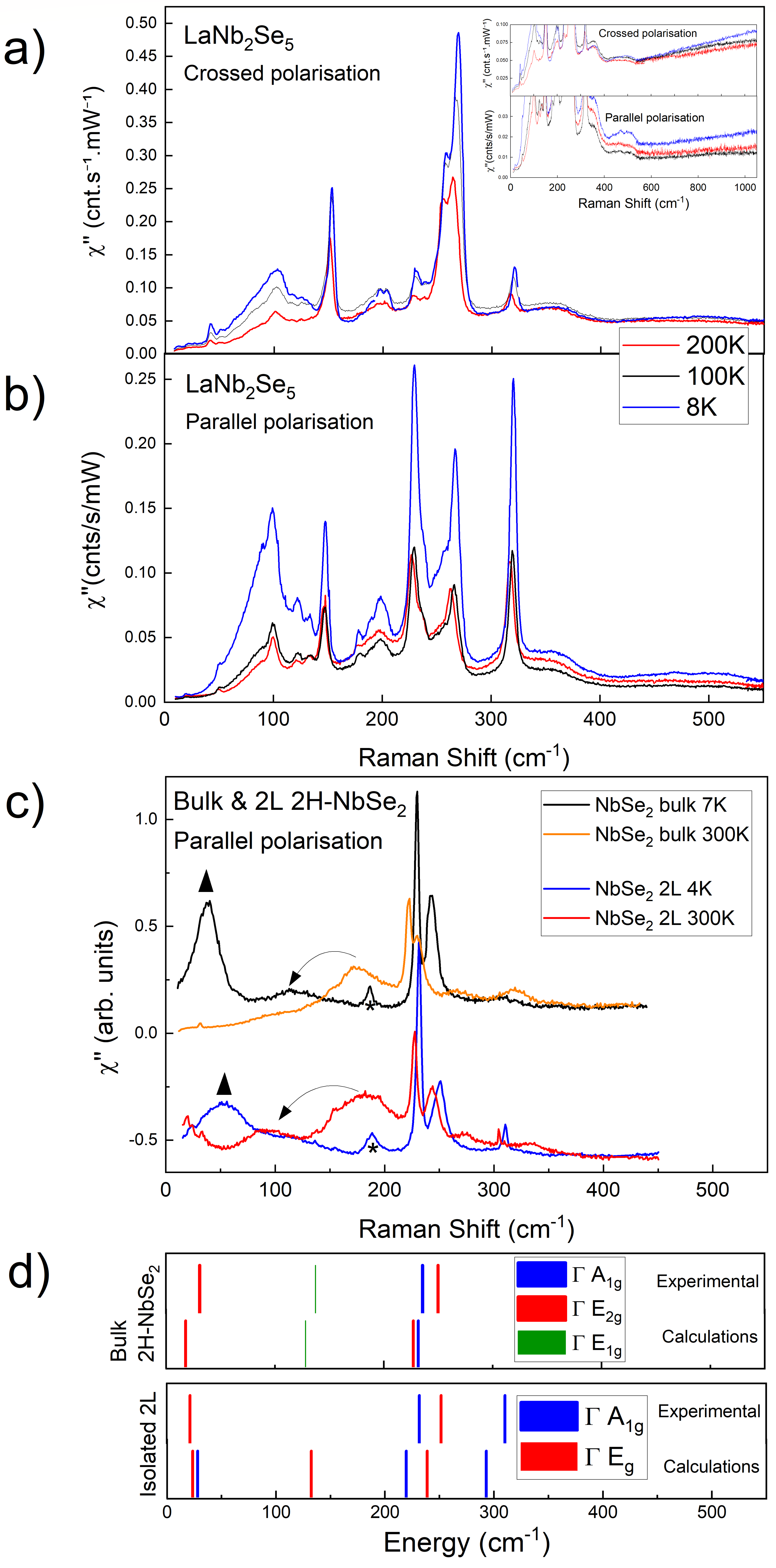}
\caption{Temperature dependence of  the Raman response of (LaSe)$_{1.14}$(NbSe$_2$)$_2$ with (ab)-plane crossed (a) and parallel (b) polarisation configurations. Inset: large energy-range electronic Raman response. c) Raman response of bulk 2H-NbSe$_2$ and isolated 2L-NbSe$_2$ \cite{Lin2020} at 300~K and $\sim$5K, in the CDW state. The stars, triangles stands for the Brillouin-zone folded phonons and amplitudons, respectively. The arrows indicate the double phonon modes related to the soft phonon branches of the CDW. (d) Raman active phonons energies from experiments and from theory.}
\label{fig3}
\end{figure}

Raman spectroscopy offers a direct probe of charge-density-wave signatures in the bulk~\cite{Sugai1981, Tsang1976}, and in few-layer systems~\cite{Xi2015}.
Two types of new Raman active modes arise as a fingerprint of the CDW.
The first one is a soft phonon called the amplitude mode, that gradually hardens when cooling down and that arises from the phonon branch which softens at the CDW wavevector. This mode has been detected in bulk $2$H-NbSe$_{2}$  at $\approx$~40~cm$^{-1}$ (triangle in Fig. \ref{fig3}, panel c) \cite{Measson2014,Grasset2018a}.  
The second type of new peaks are zone-folded modes that arise from other phonon bands at the CDW wavevector. These modes are folded into $\Gamma$ by the effect of the CDW modulation and are therefore detectable (Cf. stars in Fig. \ref{fig3}, panel c).

Figs.~\ref{fig3} (a) and (b) show the Raman response of (LaSe)$_{1.14}$(NbSe$_2$)$_2$ in crossed and parallel polarisations for temperatures ranging from $8$ to $200$ K. In both polarizations, a substantial increase of the overall intensity  is measured when cooling down. Narrow phonon modes are reported up to 350~cm$^{-1}$. The  modes above $350$ cm$^{-1}$ are broader and are most likely due to double phonon excitations.

Globally, the phonon modes harden when cooling down, as it is generally expected from anharmonic effect. No new modes appear at low temperatures, neither across the temperature range  where STM was detecting small patches with short range $2\times 2$ modulation ($100 $K) \cite{Leriche2021}, nor across the temperature range at which the CDW is detected in bulk samples ($35$ K).
The large-range electronic response does not present any signature of electronic gap opening that is sometimes measured in the CDW state \cite{Grasset2019,He2024}.

The last possible fingerprint of the presence of a CDW is a two-phonon Raman feature from the soft phonon branch at ${\bf Q}_{CDW}$, i.e. the phonon momentum related to the CDW instability. In bulk 2H-NbSe$_2$, it is visible in Fig. \ref{fig3}c) as indicated by the arrow.
The only candidate for this experimental Raman feature is the broad mode in the low energy range around $100$ cm$^{-1}$ which is detected in both polarization configurations. However, the temperature dependence of this mode is peculiar and in stark contradiction with the behaviour of the double phonon mode in NbSe$_2$.
Indeed, as shown Fig. \ref{fig3}c), in bulk $2$H-NbSe$_{2}$ the double phonon feature loses intensity in both A$_{1g}$ and E$_{2g}$ symmetries and softens with decreasing temperature.
Conversely, in the case of our (LaSe)$_{1.14}$(NbSe$_2$)$_2$, the large spectral weight bump always remains in the same energy range, and its intensity largely grows when cooling down as shown in Fig. \ref{fig3}. So even if this part of the spectra could be partially due to two phonon scattering, it does not evidence a softening of the branch and, thus, it is not related to a CDW.
A comparison with DFT calculations suggests that the nature of the broad mode in the misfit can be attributed to the presence of a dense population of LaSe modes that overlap with a few low energy NbSe$_{2}$ frequencies.    
Overall, these measurements suggest that no amplitude modes or CDW related modes occur down to $8$~K.

We comment here on the CDW signatures observed by STM at the cleaved (LaSe)$_{1.14}$(NbSe$_2$)$_2$ surface while no signature in the bulk could be detected by Raman spectroscopy. Investigation of the Raman response of 2H-NbSe$_2$ as a function of quality of the samples, as stated by the residual resistivity ratio (RRR= 50 for good samples and 6 for the worst samples), clearly shows that the main CDW signature observed by Raman spectroscopy, namely the amplitudon, becomes extremely weak intensity in low quality samples \cite{Sooryakumar1981a}.
A first hypothesis would be that the CDW would exist in the bulk but with a very short coherence length of $\approx2$~nm, as suggested by the STM experiment. In this case, the situation would be somehow analogous to the one observed in the normal state of 2H-NbSe$_2$, where it is reported that short range CDW modulations are observed by STM near the defects much above the bulk CDW critical temperature \cite{Arguello2014,Oh2020,Sahoo2022}, while no Raman signatures are detected in this temperature regime.
A second hypothesis would be that a surface peculiar behaviour would stabilize and enhance a surface CDW, while its bulk counterpart would develop at much lower temperature and with lower amplitude and coherence length, or even not form at all. There have been reports of such complicated and different surface versus bulk CDW properties in well-known quasi-one dimensional materials such as NbSe$_3$ or the blue bronze \cite{Brun2010, Machado-Charry2006,Brun2007}.

\section{Raman scattering and mode attribution}

\begin{figure}[]
\centering
\includegraphics[width=1\linewidth]{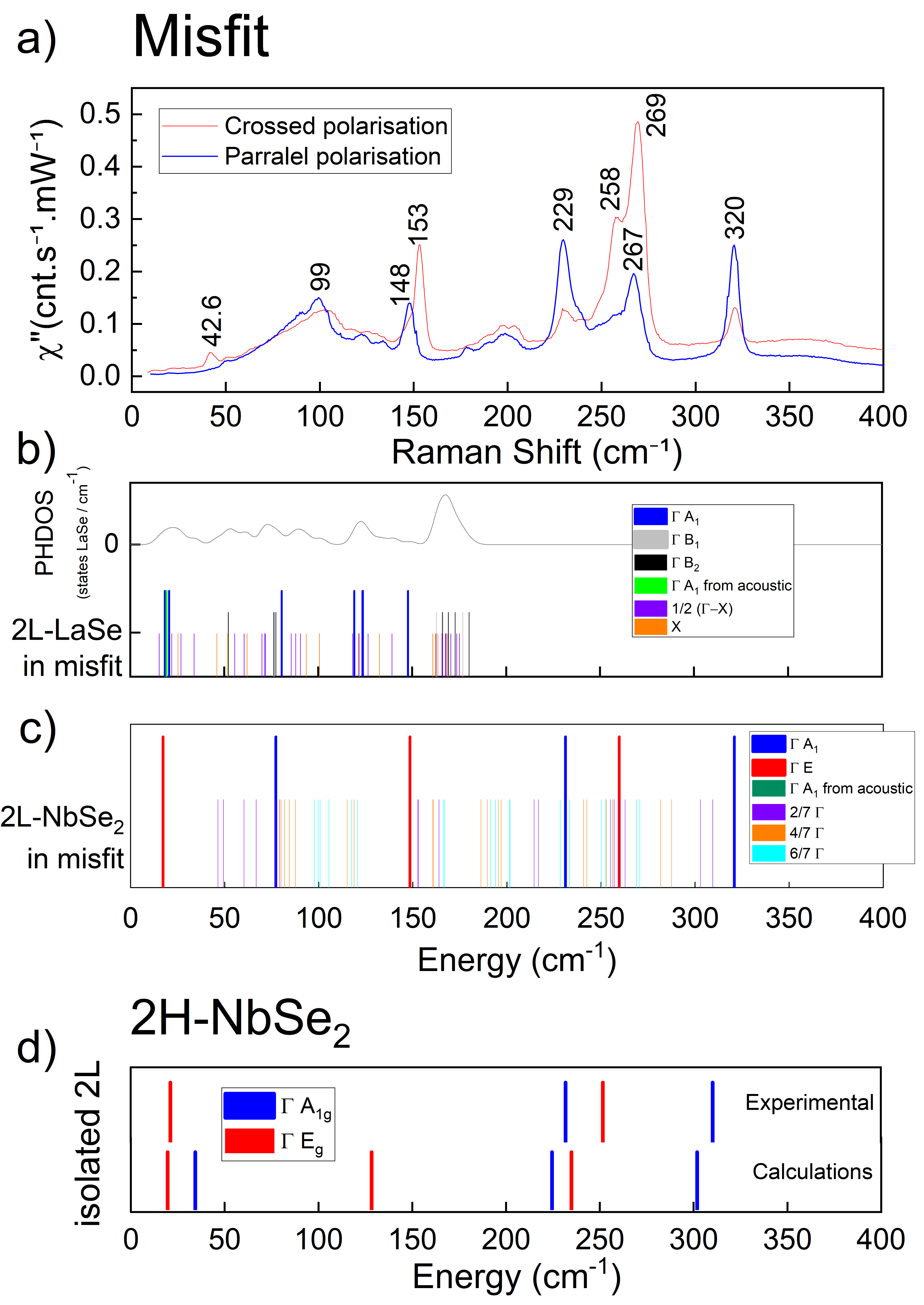}
\caption{a) Raman spectra of (LaSe)$_{1.14}$(NbSe$_2$)$_2$ at 8~K in crossed and parallel polarizations in the (ab)-plane. 
b) and c) Theoretical calculations of Raman active modes of (LaSe)$_{1.14}$(NbSe$_2$)$_2$ obtained from those of the two subsystems, namely LaSe (panel b)) and NbSe$_2$ (panel c)), electron doped by means of the FET setup. 
The original modes from $\Gamma$ of the $1\times1$ cell (long bars) and their subsequent BZ folding (short bars) modes are indicated (Cf. text). Original modes from $\Gamma$ of the $1\times1$ cell are classified by symmetries. The phonon density of states of LaSe is shown in the upper part of panel b).
d) Comparison between experimental~\cite{Lin2020} and theoretical zone center modes of isolated 2L-NbSe$_2$. Here, the NbSe$_2$ lattice parameter is the one in the misfit, namely a=$3.437$ \AA.}
\label{fig4}
\end{figure}

In Fig. \ref{fig4} we show the Raman spectra of (LaSe)$_{1.14}$(NbSe$_2$)$_2$ at 8 K in both parallel and crossed polarizations. There is a substantial difference among the two spectra, supporting a strong dependence of the signal on the symmetry of the modes. In Tab.~\ref{tab2} we report the most intense modes together with their Raman active channels.

By comparing with the experimental Raman spectra in the bulk and in the 2L-NbSe$_2$ in Fig. \ref{fig3}c) and d), we notice a striking resemblance on the symmetry of the most intense modes, especially at high energy. Notably in the 2L system, in addition to the expected 3 modes of the bulk 2H-NbSe$_2$, namely one E$_{(2)g}$ interlayer mode at $\sim$30~cm$^{-1}$, one A$_{1g}$ and one E$_{(2)g}$ mode at $\sim$~250~cm$^{-1}$, Lin et al. \cite{Lin2020} report one additional mode due to the few-layer structures, namely an A$_{1g}$ mode at high energy 310~cm$^{-1}$. A mode at $\sim$~155~cm$^{-1}$ is possibly measured by Lin et al., but would require some confirmation. Importantly, these modes, even if measured at low temperature, are not due to the CDW ordering. As shown Fig. \ref{fig3} d), the energy and symmetry of the modes are well reproduced by our calculations for both systems, with a general tendency to underestimate their energy compared to the experimental results. As a straightforward interpretation of the spectra, we then tentatively assigned the most intense modes in the misfit to the modes of the same symmetry in the 2 layer structure. As shown Table~\ref{tab2}, there is a good correspondence with 4 modes, in terms of energy and symmetry.

In order to further corroborate our analysis and perform a full assignment of the modes, we consider here the two subunits of the compounds, namely LaSe and NbSe$_2$, as well as their interplay. 
As the space group of the bulk misfit compound is P$_1$, all vibrational modes are Raman active. Thus, in the absence of a charge density wave instability, besides the zone center modes related to the LaSe and NbSe$_2$ subunit cells, one expects (i) NbSe$_2$ modes at in-plane phonon momenta that are not at zone center in the NbSe$_2$ bilayer Brillouin zone  but are  backfolded at zone center in the misfit Brillouin zone due to the $7\times 1$ NbSe$_2$ periodicity occurring in (LaSe)$_{1.14}$(NbSe$_2$)$_2$, (ii) LaSe modes at in-plane phonon momenta that are not at zone center in the LaSe bilayer Brillouin zone but are backfolded at zone center in the misfit Brillouin zone due to the  the $4\times 1$ LaSe periodicity occurring in (LaSe)$_{1.14}$(NbSe$_2$)$_2$, (iii)  modes arising from the presence of two inequivalent (LaSe)$_{1.14}$(NbSe$_2$)$_2$ units along the $c-$axis of the misfit unit cell (see Fig. \ref{fig1}) and, finally, (iv) modes that cannot be interpreted as pure LaSe or NbSe$_2$ modes.

As we will see later from PHDOS calculations of the whole misfit, practically all modes can be interpreted as modes of the two separated subunits. Thus, the occurrence of phonon modes that are mixed modes of the LaSe and NbSe$_2$ subunits can be excluded and point (iv) can be neglected. The splitting of phonon frequencies due to the presence of two inequivalent (LaSe)$_{1.14}$(NbSe$_2$)$_2$ units along the c-axis, i.e. point (iii), is also expected to be negligible as the (LaSe)$_{1.14}$(NbSe$_2$)$_2$ units are weakly interacting along the c-axis. It then follows that an attempt of interpreting the Raman response in terms of the backfolded modes of the NbSe$_2$ and LaSe subunits should lead to a clear understanding of the Raman spectra.
Thus, we proceed to a more detailed analysis based on the Raman activity of the isolated and charged LaSe and NbSe$_2$ bilayers.

The LaSe rock salt subunit (\#35, $\rm C_{2v}$ point group), has $24$ $\Gamma$-point frequencies:
$$\Gamma_{\rm LaSe}=8A_1+8B_1+8B_2$$
From symmetry, we expect B$_1$ and B$_2$ modes being Raman active, since there is no inversion center. Even so, B$_1$ and B$_2$ modes are not expected to be measured in the configuration of measurement (with Poynting vector along c axis).
On the contrary, the $8$ A$_1$ modes are Raman active and mainly in the parallel configuration since they have a $(a-b)^2$ response in crossed polarization leading to small intensities. To summarize, the only modes that can be identified in parallel polarization, expected from the pure rocksalt subsystem, have A$_1$ symmetry.

For the NbSe$_{2}$ bilayer having P$\overline{3}$m$1$ space group (\#164, D$_{3d}^{3}$ point group), the behaviour of the modes is the following. We have $18$ $\Gamma$-point frequencies as
$$\Gamma_{\rm NbSe_{2}}=3A_{1g}+3E_{g}+3A_{2u}+3E_{u}$$
The $3$ completely symmetric A$_{1g}$ modes are all Raman active only in parallel polarization. 
The $3$ double degenerate $E_{g}$ modes can be detected in both crossed and parallel polarizations.
Finally, the A$_{2u}$ and E$_{u}$ are not  Raman active.
The symmetry of the $\Gamma$ modes, as well as their activity in different polarization configurations, are reported in Tab. \ref{tab2}, in the 4$^{th}$ and 6$^{th}$ column.

\begin{table*}[t]

\resizebox{17cm}{!}{ 
\begin{threeparttable}
\begin{tabular}{|c|c|c||c|c||c|}
\hline
   &   &   &  &  & \\
 \,\,Intense Modes\,\,  & \,\,Raman activity\,\,  & Experimental & \,\,Calculated modes\,\,   &    \,\,Calculated modes\,\,  &    \,\,Calculated modes\,\,  \\ %

of (LaSe)$_{1.14}$(NbSe$_2$)$_2$& $\perp$ or $\parallel$  & modes & energy/symmetry & from splitted E$_g$ & energy/symmetry  \\

(in cm$^{-1}$)    &  &   in 2L NbSe$_2$    & \,from 2L-NbSe$_2$ at $\Gamma$ \, & \, in misfit \, & from LaSe at $\Gamma$\\ 
   &   &   &  & & \\
\hline
   &   &   &  & & \\
42.6 & $\perp$& 21/E$_g$ &  17.3/E$_g$  & & \\

99.0& $\parallel$ &  &  77.2/A$_{1g}$ & & or 120.1/A$_{1}$ \\

148&$\parallel$&  &  & & 148.9/A$_{1}$ \\

153 & $\perp$ & 154 \tnote{a}  & 148.5/E$_g$ & & \\

229 & $\parallel$ & 232/A$_{1g}$ & 231.2/A$_{1g}$ & &\\

258 & $\perp$ & 251.5/E$_g$ &  \textcolor{blue}{259.7/E$_g$}& \textcolor{blue}{251.5/E$_g$}& \\

267 & $\parallel$ & & X & & X\\

269 & $\perp$ & & &  \textcolor{blue}{258.7/E$_g$} &\\
320 & $\parallel$ & 310.5/A$_{1g}$ &  320.9/A$_{1g}$ & & \\
   &   &   &   &  & \\                     
\hline 

\end{tabular}
\begin{tablenotes}
\item[a] Xi Xiaoxiang, private communication: This mode at 154~cm$^{-1}$ may require experimental confirmation.
\end{tablenotes}
\end{threeparttable}
}
\caption{Intense Raman active modes measured in (LaSe)$_{1.14}$(NbSe$_2$)$_2$ at 8~K with polarizations in the (ab) plane. 
The labels $\perp$ and $\parallel$ stand for crossed and parallel polarizations, respectively.  The 3$^{rd}$ column reports the experimental data in an isolated 2L-NbSe$_2$ \cite{Lin2020}. Calculated phonon modes in the misfit structure originating from the $\Gamma$ point of the unfolded Brillouin Zone and from subunits of NbSe$_2$ and LaSe in the 4$^{th}$ and 6$^{th}$ columns, respectively. In the 5$^{th}$ column, the splitted E$_g$ modes of the $8/4$ misfit cell which correspond to the former E$_{g}$ of NbSe$_2$ subunit (in blue). The label X marks the single intense mode that we could not assign from the two subunits (Cf. text).}
\label{tab2}
\end{table*}

In the table, we assign the calculated $\Gamma$ frequencies to the most intense mode in the Raman spectra.
As it can be seen, only one of the most intense modes can be ascribed to LaSe. A second one at 99~cm$^{-1}$ could be either assigned to LaSe or NbSe$_2$ since both subunits present a parallel-active mode in this range of energy. The other peaks are all derived from the $2$L-NbSe$_{2}$ subsystem. The physics of the Raman spectra at 8~K reveals that the lattice dynamics of (LaSe)$_{1.14}$(NbSe$_2$)$_2$ can be described in terms of that of its individual constituents.

\begin{figure}[]
\centering
\includegraphics[width=1\linewidth]{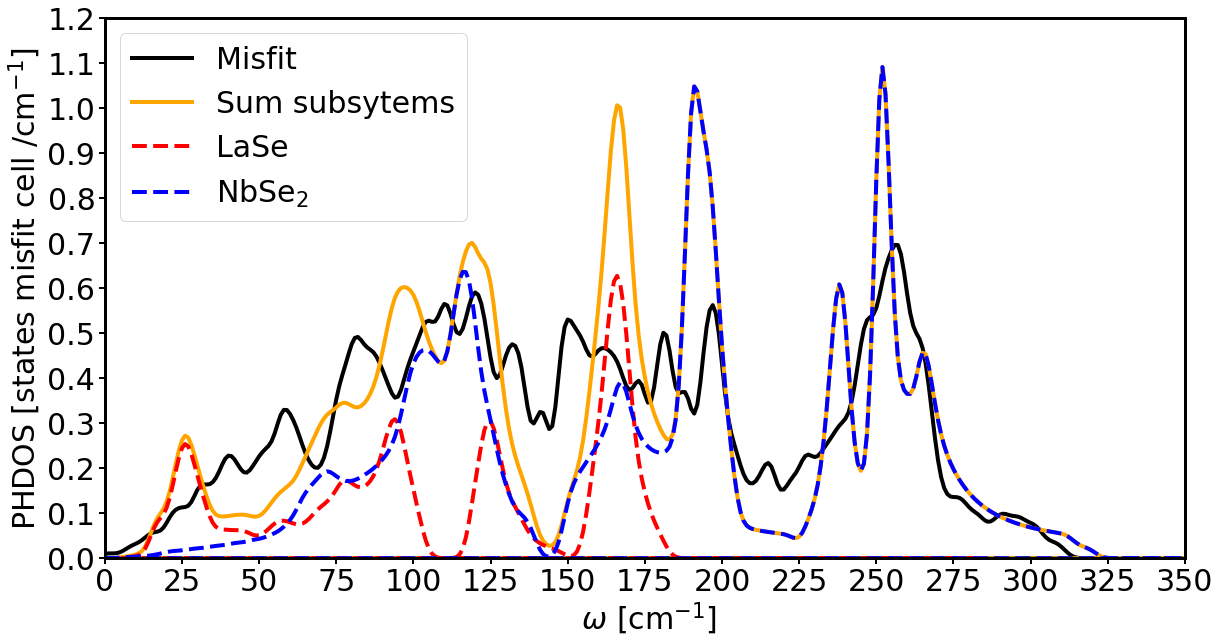}
\caption{ Phonon density of states (PHDOS) per misfit formula unit of bulk (LaSe)$_{1.14}$(NbSe$_2$)$_2$
(black solid line). Comparison with the PHDOS of the individual subsystems: (red dashed line) LaSe subunit doped of $1.2$ electrons per unit cell in the FET setup, (blue dashed line) NbSe$_{2}$ bilayer doped of $0.6$ electrons per Nb atom in the FET setup, respectively, and with their sum (orange solid line).}
\label{fig5}
\end{figure}

To strengthen this statement, we can look at the PHDOS calculation on the $8/4=2/1$ periodic approximant of the full misfit is shown in Fig. \ref{fig5}. The results are compared in Fig. \ref{fig5} with the phonon density of states of the two separated subunits as well as with their sum in the presence of a field effect charging mimicking the charge transfer among the LaSe and NbSe$_2$ subunits. As depicted in Fig. \ref{fig5}, almost all features in the misfit PHDOS are fairly well explained in terms of the sum of the PHDOS of the two (field-effect charged) separated subunits. The only feature present in the misfit PHDOS, but not in the PHDOS of the two subunits, is a peak at $\approx 150$ cm$^{-1}$. This peak is at slightly higher energies $\approx 165-170$ cm$^{-1}$ in the LaSe subunit. The difference is  due to the strain applied to the LaSe subunit inside the misfit to obtain the $2/1$ periodic approximant ($14\%$ strain), while the field-effect transistor (FET) charged LaSe bilayer is unstrained and has the same lattice parameters as in the bulk misfit. Overall, we can  state that the vibrational properties of the (LaSe)$_{1.14}$(NbSe$_2$)$_2$ are entirely determined by those of the two separated subunits with an appropriate amount of charging.

From Fig. \ref{fig5} it is also clear that, due to the heavy La mass, the phonon modes of the RS subunit are mostly concentrated in the low energy part of the spectrum (below $175$ cm$^{-1}$), while those of the NbSe$_2$ bilayer occurs at all energies.

As summarized in Tab. \ref{tab2} only two of the most intense Raman peak, namely the one at $269$ cm$^{-1}$ in crossed polarization and $267$ cm$^{-1}$ in parallel polarization, are not directly deducible from $2$L-NbSe$_{2}$'s modes. 
We are able to assign the highly intense mode at $269$ cm$^{-1}$ in crossed polarization to the former double-degenerate high energy E$_g$ mode of $2$L-NbSe$_{2}$ at $259$ cm$^{-1}$ that splits in the misfit.

In order to perform this assignment we consider the full misfit calculation employing the $8/4=2/1$ periodic approximant (\ref{fig5}), where the two distinct peaks can be clearly identified in this energy range. In order to check if these peaks originate from the E$_g$ mode of the isolated FET-doped $2$L-NbSe$_{2}$, we project all the full misfit phonon eigenvectors onto the ones corresponding to the doubly degenerate E$_g$ mode at $259$ cm$^{-1}$ in the isoated FET-doped $2$L-NbSe$_{2}$. We find that the highest E$_g$ character is present in two modes at $251.5$ and $258.7$ cm$^{-1}$ (see Tab. \ref{tab2}, 5th column).

Note that we also evaluated the effect of the slight non-hexagonality of $2$L-NbSe$_{2}$ within the misfit \cite{Leriche2021}, which is just a very small $\sim~1$ cm$^{-1}$ splitting that cannot account for our experimental results (Cf. SM).

Finally, only the intense mode at $267$ cm$^{-1}$ in parallel polarization is not captured by our DFT calculations. This one is most probably a hybrid mode of the system as a whole, caused by the bonding between the TMD and the RS subunits that we neglected in our calculations.

\section{Conclusions}

By using polarized Raman spectroscopy and first principles calculations, we provide a complete description of the vibrational properties of the misfit layer compound (LaSe)$_{1.14}$(NbSe$_2$)$_2$. 
We identify all the main phonon modes and their symmetry and demonstrate that, similarly to what happens for the electronic properties, the vibrational properties can be understood in terms of the two subunits (LaSe and NbSe$_2$ bilayers) in a field effect configuration, where the charging of the gates is directly determined by the charge transfer in the misfit structure. Notably, the lattice dynamics of the TMD has a strong 2D character in this 3D misfit structure. 
Finally, our theoretical understanding is supported by the Raman results, particularly by the charge density wave collapse in the misfit due to the large charge transfer from the LaSe subunit. Our work is relevant beyond the case of (LaSe)$_{1.14}$(NbSe$_2$)$_2$ as it sets a roadmap for the investigation of the large class of materials composed by misfit layer compounds.

\section*{Acknowledgements}
We thank Xiaoxiang Xi for fruitful exchanges of information. We thank Alex W. Chin for fruitful discussions. M.-A.M. thanks the European Research Council (ERC) under the European Union's Horizon 2020 research and innovation programme (Grant Agreement n$^{\circ}$ 865826). This work has received funding from the Agence Nationale de la Recherche under the project Misfit (Projet-ANR-21-CE30-0054). M.C. acknowledges support from ICSC – Centro Nazionale di Ricerca in HPC, Big Data and Quantum Computing, funded by the European Union under NextGenerationEU. Views and opinions expressed are however those of the author(s) only and do not necessarily reflect those of the European Union or The European Research Executive Agency.  Neither the European Union nor the granting authority can be held responsible for them.

\bibliography{apssamp}

\begin{thebibliography}{42}%
\makeatletter
\providecommand \@ifxundefined [1]{%
 \@ifx{#1\undefined}
}%
\providecommand \@ifnum [1]{%
 \ifnum #1\expandafter \@firstoftwo
 \else \expandafter \@secondoftwo
 \fi
}%
\providecommand \@ifx [1]{%
 \ifx #1\expandafter \@firstoftwo
 \else \expandafter \@secondoftwo
 \fi
}%
\providecommand \natexlab [1]{#1}%
\providecommand \enquote  [1]{``#1''}%
\providecommand \bibnamefont  [1]{#1}%
\providecommand \bibfnamefont [1]{#1}%
\providecommand \citenamefont [1]{#1}%
\providecommand \href@noop [0]{\@secondoftwo}%
\providecommand \href [0]{\begingroup \@sanitize@url \@href}%
\providecommand \@href[1]{\@@startlink{#1}\@@href}%
\providecommand \@@href[1]{\endgroup#1\@@endlink}%
\providecommand \@sanitize@url [0]{\catcode `\\12\catcode `\$12\catcode
  `\&12\catcode `\#12\catcode `\^12\catcode `\_12\catcode `\%12\relax}%
\providecommand \@@startlink[1]{}%
\providecommand \@@endlink[0]{}%
\providecommand \url  [0]{\begingroup\@sanitize@url \@url }%
\providecommand \@url [1]{\endgroup\@href {#1}{\urlprefix }}%
\providecommand \urlprefix  [0]{URL }%
\providecommand \Eprint [0]{\href }%
\providecommand \doibase [0]{https://doi.org/}%
\providecommand \selectlanguage [0]{\@gobble}%
\providecommand \bibinfo  [0]{\@secondoftwo}%
\providecommand \bibfield  [0]{\@secondoftwo}%
\providecommand \translation [1]{[#1]}%
\providecommand \BibitemOpen [0]{}%
\providecommand \bibitemStop [0]{}%
\providecommand \bibitemNoStop [0]{.\EOS\space}%
\providecommand \EOS [0]{\spacefactor3000\relax}%
\providecommand \BibitemShut  [1]{\csname bibitem#1\endcsname}%
\let\auto@bib@innerbib\@empty
\bibitem [{\citenamefont {Manzeli}\ \emph {et~al.}(2017)\citenamefont
  {Manzeli}, \citenamefont {Ovchinnikov}, \citenamefont {Pasquier},
  \citenamefont {Yazyev},\ and\ \citenamefont {Kis}}]{MANZELI2017}%
  \BibitemOpen
  \bibfield  {author} {\bibinfo {author} {\bibfnamefont {S.}~\bibnamefont
  {Manzeli}}, \bibinfo {author} {\bibfnamefont {D.}~\bibnamefont
  {Ovchinnikov}}, \bibinfo {author} {\bibfnamefont {D.}~\bibnamefont
  {Pasquier}}, \bibinfo {author} {\bibfnamefont {O.~V.}\ \bibnamefont
  {Yazyev}},\ and\ \bibinfo {author} {\bibfnamefont {A.}~\bibnamefont {Kis}},\
  }\bibfield  {title} {\bibinfo {title} {{{2D}} transition metal
  dichalcogenides},\ }\href {https://doi.org/10.1038/natrevmats.2017.33}
  {\bibfield  {journal} {\bibinfo  {journal} {Nature Reviews Materials}\
  }\textbf {\bibinfo {volume} {2}},\ \bibinfo {pages} {17033} (\bibinfo {year}
  {2017})}\BibitemShut {NoStop}%
\bibitem [{\citenamefont {Fu}\ \emph {et~al.}(2021)\citenamefont {Fu},
  \citenamefont {Han}, \citenamefont {Wang}, \citenamefont {Xu}, \citenamefont
  {Yao}, \citenamefont {Zhong}, \citenamefont {Zhong}, \citenamefont {Liu},
  \citenamefont {Gao}, \citenamefont {Zhang}, \citenamefont {Xu},\ and\
  \citenamefont {Song}}]{FU2021}%
  \BibitemOpen
  \bibfield  {author} {\bibinfo {author} {\bibfnamefont {Q.}~\bibnamefont
  {Fu}}, \bibinfo {author} {\bibfnamefont {J.}~\bibnamefont {Han}}, \bibinfo
  {author} {\bibfnamefont {X.}~\bibnamefont {Wang}}, \bibinfo {author}
  {\bibfnamefont {P.}~\bibnamefont {Xu}}, \bibinfo {author} {\bibfnamefont
  {T.}~\bibnamefont {Yao}}, \bibinfo {author} {\bibfnamefont {J.}~\bibnamefont
  {Zhong}}, \bibinfo {author} {\bibfnamefont {W.}~\bibnamefont {Zhong}},
  \bibinfo {author} {\bibfnamefont {S.}~\bibnamefont {Liu}}, \bibinfo {author}
  {\bibfnamefont {T.}~\bibnamefont {Gao}}, \bibinfo {author} {\bibfnamefont
  {Z.}~\bibnamefont {Zhang}}, \bibinfo {author} {\bibfnamefont
  {L.}~\bibnamefont {Xu}},\ and\ \bibinfo {author} {\bibfnamefont
  {B.}~\bibnamefont {Song}},\ }\bibfield  {title} {\bibinfo {title} {{{2D
  Transition Metal Dichalcogenides}}: {{Design}}, {{Modulation}}, and
  {{Challenges}} in {{Electrocatalysis}}},\ }\href
  {https://doi.org/10.1002/adma.201907818} {\bibfield  {journal} {\bibinfo
  {journal} {Advanced Materials}\ }\textbf {\bibinfo {volume} {33}},\ \bibinfo
  {pages} {1907818} (\bibinfo {year} {2021})}\BibitemShut {NoStop}%
\bibitem [{\citenamefont {Chen}\ \emph {et~al.}(2022)\citenamefont {Chen},
  \citenamefont {Zhang}, \citenamefont {Kan}, \citenamefont {He}, \citenamefont
  {Song}, \citenamefont {Pang}, \citenamefont {Wei},\ and\ \citenamefont
  {Chen}}]{CHEN2022}%
  \BibitemOpen
  \bibfield  {author} {\bibinfo {author} {\bibfnamefont {H.}~\bibnamefont
  {Chen}}, \bibinfo {author} {\bibfnamefont {J.}~\bibnamefont {Zhang}},
  \bibinfo {author} {\bibfnamefont {D.}~\bibnamefont {Kan}}, \bibinfo {author}
  {\bibfnamefont {J.}~\bibnamefont {He}}, \bibinfo {author} {\bibfnamefont
  {M.}~\bibnamefont {Song}}, \bibinfo {author} {\bibfnamefont {J.}~\bibnamefont
  {Pang}}, \bibinfo {author} {\bibfnamefont {S.}~\bibnamefont {Wei}},\ and\
  \bibinfo {author} {\bibfnamefont {K.}~\bibnamefont {Chen}},\ }\bibfield
  {title} {\bibinfo {title} {The {{Recent Progress}} of {{Two-Dimensional
  Transition Metal Dichalcogenides}} and {{Their Phase Transition}}},\ }\href
  {https://doi.org/10.3390/cryst12101381} {\bibfield  {journal} {\bibinfo
  {journal} {Crystals}\ }\textbf {\bibinfo {volume} {12}},\ \bibinfo {pages}
  {1381} (\bibinfo {year} {2022})}\BibitemShut {NoStop}%
\bibitem [{\citenamefont {Joseph}\ \emph {et~al.}(2023)\citenamefont {Joseph},
  \citenamefont {Mohan}, \citenamefont {Lakshmy}, \citenamefont {Thomas},
  \citenamefont {Chakraborty}, \citenamefont {Thomas},\ and\ \citenamefont
  {Kalarikkal}}]{JOSEPH2023}%
  \BibitemOpen
  \bibfield  {author} {\bibinfo {author} {\bibfnamefont {S.}~\bibnamefont
  {Joseph}}, \bibinfo {author} {\bibfnamefont {J.}~\bibnamefont {Mohan}},
  \bibinfo {author} {\bibfnamefont {S.}~\bibnamefont {Lakshmy}}, \bibinfo
  {author} {\bibfnamefont {S.}~\bibnamefont {Thomas}}, \bibinfo {author}
  {\bibfnamefont {B.}~\bibnamefont {Chakraborty}}, \bibinfo {author}
  {\bibfnamefont {S.}~\bibnamefont {Thomas}},\ and\ \bibinfo {author}
  {\bibfnamefont {N.}~\bibnamefont {Kalarikkal}},\ }\bibfield  {title}
  {\bibinfo {title} {A review of the synthesis, properties, and applications of
  {{2D}} transition metal dichalcogenides and their heterostructures},\ }\href
  {https://doi.org/10.1016/j.matchemphys.2023.127332} {\bibfield  {journal}
  {\bibinfo  {journal} {Materials Chemistry and Physics}\ }\textbf {\bibinfo
  {volume} {297}},\ \bibinfo {pages} {127332} (\bibinfo {year}
  {2023})}\BibitemShut {NoStop}%
\bibitem [{\citenamefont {Wilson}\ \emph {et~al.}(1975)\citenamefont {Wilson},
  \citenamefont {Di~Salvo},\ and\ \citenamefont {Mahajan}}]{Wilson1975}%
  \BibitemOpen
  \bibfield  {author} {\bibinfo {author} {\bibfnamefont {J.}~\bibnamefont
  {Wilson}}, \bibinfo {author} {\bibfnamefont {F.}~\bibnamefont {Di~Salvo}},\
  and\ \bibinfo {author} {\bibfnamefont {S.}~\bibnamefont {Mahajan}},\
  }\bibfield  {title} {\bibinfo {title} {Charge-density waves and superlattices
  in the metallic layered transition metal dichalcogenides},\ }\href
  {https://doi.org/10.1080/00018737500101391} {\bibfield  {journal} {\bibinfo
  {journal} {Advances in Physics}\ }\textbf {\bibinfo {volume} {24}},\ \bibinfo
  {pages} {117} (\bibinfo {year} {1975})}\BibitemShut {NoStop}%
\bibitem [{\citenamefont {Moncton}\ \emph {et~al.}(1977)\citenamefont
  {Moncton}, \citenamefont {Axe},\ and\ \citenamefont
  {DiSalvo}}]{Moncton1977a}%
  \BibitemOpen
  \bibfield  {author} {\bibinfo {author} {\bibfnamefont {D.~E.}\ \bibnamefont
  {Moncton}}, \bibinfo {author} {\bibfnamefont {J.~D.}\ \bibnamefont {Axe}},\
  and\ \bibinfo {author} {\bibfnamefont {F.~J.}\ \bibnamefont {DiSalvo}},\
  }\bibfield  {title} {\bibinfo {title} {Neutron scattering study of the
  charge-density wave transitions in 2{{H}}-{{TaSe}}$_{2}$ and
  2{{H}}-{{NbSe}}$_{2}$},\ }\href {https://doi.org/10.1103/PhysRevB.16.801}
  {\bibfield  {journal} {\bibinfo  {journal} {Physical Review B}\ }\textbf
  {\bibinfo {volume} {16}},\ \bibinfo {pages} {801} (\bibinfo {year}
  {1977})}\BibitemShut {NoStop}%
\bibitem [{\citenamefont {Malliakas}\ and\ \citenamefont
  {Kanatzidis}(2013)}]{Malliakas2013}%
  \BibitemOpen
  \bibfield  {author} {\bibinfo {author} {\bibfnamefont {C.~D.}\ \bibnamefont
  {Malliakas}}\ and\ \bibinfo {author} {\bibfnamefont {M.~G.}\ \bibnamefont
  {Kanatzidis}},\ }\bibfield  {title} {\bibinfo {title} {Nb--{{Nb Interactions
  Define}} the {{Charge Density Wave Structure}} of {{2H-NbSe}}
  {\textsubscript{2}}},\ }\href {https://doi.org/10.1021/ja3120554} {\bibfield
  {journal} {\bibinfo  {journal} {Journal of the American Chemical Society}\
  }\textbf {\bibinfo {volume} {135}},\ \bibinfo {pages} {1719} (\bibinfo {year}
  {2013})}\BibitemShut {NoStop}%
\bibitem [{\citenamefont {Revolinsky}\ \emph {et~al.}(1965)\citenamefont
  {Revolinsky}, \citenamefont {Spiering},\ and\ \citenamefont
  {Beerntsen}}]{Revolinsky1965}%
  \BibitemOpen
  \bibfield  {author} {\bibinfo {author} {\bibfnamefont {E.}~\bibnamefont
  {Revolinsky}}, \bibinfo {author} {\bibfnamefont {G.}~\bibnamefont
  {Spiering}},\ and\ \bibinfo {author} {\bibfnamefont {D.}~\bibnamefont
  {Beerntsen}},\ }\bibfield  {title} {\bibinfo {title} {Superconductivity in
  the niobium-selenium system},\ }\href
  {https://doi.org/10.1016/0022-3697(65)90190-3} {\bibfield  {journal}
  {\bibinfo  {journal} {Journal of Physics and Chemistry of Solids}\ }\textbf
  {\bibinfo {volume} {26}},\ \bibinfo {pages} {1029} (\bibinfo {year}
  {1965})}\BibitemShut {NoStop}%
\bibitem [{\citenamefont {Xi}\ \emph {et~al.}(2015)\citenamefont {Xi},
  \citenamefont {Zhao}, \citenamefont {Wang}, \citenamefont {Berger},
  \citenamefont {Forr{\'o}}, \citenamefont {Shan},\ and\ \citenamefont
  {Mak}}]{Xi2015}%
  \BibitemOpen
  \bibfield  {author} {\bibinfo {author} {\bibfnamefont {X.}~\bibnamefont
  {Xi}}, \bibinfo {author} {\bibfnamefont {L.}~\bibnamefont {Zhao}}, \bibinfo
  {author} {\bibfnamefont {Z.}~\bibnamefont {Wang}}, \bibinfo {author}
  {\bibfnamefont {H.}~\bibnamefont {Berger}}, \bibinfo {author} {\bibfnamefont
  {L.}~\bibnamefont {Forr{\'o}}}, \bibinfo {author} {\bibfnamefont
  {J.}~\bibnamefont {Shan}},\ and\ \bibinfo {author} {\bibfnamefont {K.~F.}\
  \bibnamefont {Mak}},\ }\bibfield  {title} {\bibinfo {title} {Strongly
  enhanced charge-density-wave order in monolayer {{NbSe$_{2}$}}},\ }\href
  {https://doi.org/10.1038/nnano.2015.143} {\bibfield  {journal} {\bibinfo
  {journal} {Nature Nanotechnology}\ }\textbf {\bibinfo {volume} {10}},\
  \bibinfo {pages} {765} (\bibinfo {year} {2015})}\BibitemShut {NoStop}%
\bibitem [{\citenamefont {Ugeda}\ \emph {et~al.}(2016)\citenamefont {Ugeda},
  \citenamefont {Bradley}, \citenamefont {Zhang}, \citenamefont {Onishi},
  \citenamefont {Chen}, \citenamefont {Ruan}, \citenamefont
  {{Ojeda-Aristizabal}}, \citenamefont {Ryu}, \citenamefont {Edmonds},
  \citenamefont {Tsai}, \citenamefont {Riss}, \citenamefont {Mo}, \citenamefont
  {Lee}, \citenamefont {Zettl}, \citenamefont {Hussain}, \citenamefont {Shen},\
  and\ \citenamefont {Crommie}}]{Ugeda2016}%
  \BibitemOpen
  \bibfield  {author} {\bibinfo {author} {\bibfnamefont {M.~M.}\ \bibnamefont
  {Ugeda}}, \bibinfo {author} {\bibfnamefont {A.~J.}\ \bibnamefont {Bradley}},
  \bibinfo {author} {\bibfnamefont {Y.}~\bibnamefont {Zhang}}, \bibinfo
  {author} {\bibfnamefont {S.}~\bibnamefont {Onishi}}, \bibinfo {author}
  {\bibfnamefont {Y.}~\bibnamefont {Chen}}, \bibinfo {author} {\bibfnamefont
  {W.}~\bibnamefont {Ruan}}, \bibinfo {author} {\bibfnamefont {C.}~\bibnamefont
  {{Ojeda-Aristizabal}}}, \bibinfo {author} {\bibfnamefont {H.}~\bibnamefont
  {Ryu}}, \bibinfo {author} {\bibfnamefont {M.~T.}\ \bibnamefont {Edmonds}},
  \bibinfo {author} {\bibfnamefont {H.-Z.}\ \bibnamefont {Tsai}}, \bibinfo
  {author} {\bibfnamefont {A.}~\bibnamefont {Riss}}, \bibinfo {author}
  {\bibfnamefont {S.-K.}\ \bibnamefont {Mo}}, \bibinfo {author} {\bibfnamefont
  {D.}~\bibnamefont {Lee}}, \bibinfo {author} {\bibfnamefont {A.}~\bibnamefont
  {Zettl}}, \bibinfo {author} {\bibfnamefont {Z.}~\bibnamefont {Hussain}},
  \bibinfo {author} {\bibfnamefont {Z.-X.}\ \bibnamefont {Shen}},\ and\
  \bibinfo {author} {\bibfnamefont {M.~F.}\ \bibnamefont {Crommie}},\
  }\bibfield  {title} {\bibinfo {title} {Characterization of collective ground
  states in single-layer {{NbSe$_{2}$}}},\ }\href
  {https://doi.org/10.1038/nphys3527} {\bibfield  {journal} {\bibinfo
  {journal} {Nature Physics}\ }\textbf {\bibinfo {volume} {12}},\ \bibinfo
  {pages} {92} (\bibinfo {year} {2016})}\BibitemShut {NoStop}%
\bibitem [{\citenamefont {Moulding}\ \emph {et~al.}(2020)\citenamefont
  {Moulding}, \citenamefont {Osmond}, \citenamefont {Flicker}, \citenamefont
  {Muramatsu},\ and\ \citenamefont {Friedemann}}]{Moulding2020}%
  \BibitemOpen
  \bibfield  {author} {\bibinfo {author} {\bibfnamefont {O.}~\bibnamefont
  {Moulding}}, \bibinfo {author} {\bibfnamefont {I.}~\bibnamefont {Osmond}},
  \bibinfo {author} {\bibfnamefont {F.}~\bibnamefont {Flicker}}, \bibinfo
  {author} {\bibfnamefont {T.}~\bibnamefont {Muramatsu}},\ and\ \bibinfo
  {author} {\bibfnamefont {S.}~\bibnamefont {Friedemann}},\ }\bibfield  {title}
  {\bibinfo {title} {Absence of superconducting dome at the charge-density-wave
  quantum phase transition in 2 {{H}}-{{NbSe}}$_{2}$},\ }\href
  {https://doi.org/10.1103/PhysRevResearch.2.043392} {\bibfield  {journal}
  {\bibinfo  {journal} {Physical Review Research}\ }\textbf {\bibinfo {volume}
  {2}},\ \bibinfo {pages} {043392} (\bibinfo {year} {2020})}\BibitemShut
  {NoStop}%
\bibitem [{\citenamefont {Wang}\ \emph {et~al.}(2020)\citenamefont {Wang},
  \citenamefont {Li}, \citenamefont {Su},\ and\ \citenamefont
  {Loh}}]{Wang2020}%
  \BibitemOpen
  \bibfield  {author} {\bibinfo {author} {\bibfnamefont {Z.}~\bibnamefont
  {Wang}}, \bibinfo {author} {\bibfnamefont {R.}~\bibnamefont {Li}}, \bibinfo
  {author} {\bibfnamefont {C.}~\bibnamefont {Su}},\ and\ \bibinfo {author}
  {\bibfnamefont {K.~P.}\ \bibnamefont {Loh}},\ }\bibfield  {title} {\bibinfo
  {title} {Intercalated phases of transition metal dichalcogenides},\ }\href
  {https://doi.org/10.1002/smm2.1013} {\bibfield  {journal} {\bibinfo
  {journal} {SmartMat}\ }\textbf {\bibinfo {volume} {1}},\ \bibinfo {pages}
  {e1013} (\bibinfo {year} {2020})}\BibitemShut {NoStop}%
\bibitem [{\citenamefont {Novoselov}\ \emph {et~al.}(2005)\citenamefont
  {Novoselov}, \citenamefont {Jiang}, \citenamefont {Schedin}, \citenamefont
  {Booth}, \citenamefont {Khotkevich}, \citenamefont {Morozov},\ and\
  \citenamefont {Geim}}]{Novoselov2005a}%
  \BibitemOpen
  \bibfield  {author} {\bibinfo {author} {\bibfnamefont {K.~S.}\ \bibnamefont
  {Novoselov}}, \bibinfo {author} {\bibfnamefont {D.}~\bibnamefont {Jiang}},
  \bibinfo {author} {\bibfnamefont {F.}~\bibnamefont {Schedin}}, \bibinfo
  {author} {\bibfnamefont {T.~J.}\ \bibnamefont {Booth}}, \bibinfo {author}
  {\bibfnamefont {V.~V.}\ \bibnamefont {Khotkevich}}, \bibinfo {author}
  {\bibfnamefont {S.~V.}\ \bibnamefont {Morozov}},\ and\ \bibinfo {author}
  {\bibfnamefont {A.~K.}\ \bibnamefont {Geim}},\ }\bibfield  {title} {\bibinfo
  {title} {Two-dimensional atomic crystals},\ }\href
  {https://doi.org/10.1073/pnas.0502848102} {\bibfield  {journal} {\bibinfo
  {journal} {Proceedings of the National Academy of Sciences}\ }\textbf
  {\bibinfo {volume} {102}},\ \bibinfo {pages} {10451} (\bibinfo {year}
  {2005})}\BibitemShut {NoStop}%
\bibitem [{\citenamefont {Xi}\ \emph {et~al.}(2016)\citenamefont {Xi},
  \citenamefont {Berger}, \citenamefont {Forr{\'o}}, \citenamefont {Shan},\
  and\ \citenamefont {Mak}}]{Xi2016b}%
  \BibitemOpen
  \bibfield  {author} {\bibinfo {author} {\bibfnamefont {X.}~\bibnamefont
  {Xi}}, \bibinfo {author} {\bibfnamefont {H.}~\bibnamefont {Berger}}, \bibinfo
  {author} {\bibfnamefont {L.}~\bibnamefont {Forr{\'o}}}, \bibinfo {author}
  {\bibfnamefont {J.}~\bibnamefont {Shan}},\ and\ \bibinfo {author}
  {\bibfnamefont {K.~F.}\ \bibnamefont {Mak}},\ }\bibfield  {title} {\bibinfo
  {title} {Gate {{Tuning}} of {{Electronic Phase Transitions}} in
  {{Two-Dimensional NbSe}}$_{2}$},\ }\href
  {https://doi.org/10.1103/PhysRevLett.117.106801} {\bibfield  {journal}
  {\bibinfo  {journal} {Physical Review Letters}\ }\textbf {\bibinfo {volume}
  {117}},\ \bibinfo {pages} {106801} (\bibinfo {year} {2016})}\BibitemShut
  {NoStop}%
\bibitem [{\citenamefont {Rouxel}\ \emph {et~al.}(1995)\citenamefont {Rouxel},
  \citenamefont {Meerschaut},\ and\ \citenamefont {Wiegers}}]{ROUXEL1995}%
  \BibitemOpen
  \bibfield  {author} {\bibinfo {author} {\bibfnamefont {J.}~\bibnamefont
  {Rouxel}}, \bibinfo {author} {\bibfnamefont {A.}~\bibnamefont {Meerschaut}},\
  and\ \bibinfo {author} {\bibfnamefont {G.}~\bibnamefont {Wiegers}},\
  }\bibfield  {title} {\bibinfo {title} {Chalcogenide misfit layer compounds},\
  }\href {https://doi.org/10.1016/0925-8388(95)01680-5} {\bibfield  {journal}
  {\bibinfo  {journal} {Journal of Alloys and Compounds}\ }\textbf {\bibinfo
  {volume} {229}},\ \bibinfo {pages} {144} (\bibinfo {year}
  {1995})}\BibitemShut {NoStop}%
\bibitem [{\citenamefont {Wiegers}(1996)}]{WIEGERS1996}%
  \BibitemOpen
  \bibfield  {author} {\bibinfo {author} {\bibfnamefont {G.}~\bibnamefont
  {Wiegers}},\ }\bibfield  {title} {\bibinfo {title} {Misfit layer compounds:
  {{Structures}} and physical properties},\ }\href
  {https://doi.org/10.1016/0079-6786(95)00007-0} {\bibfield  {journal}
  {\bibinfo  {journal} {Progress in Solid State Chemistry}\ }\textbf {\bibinfo
  {volume} {24}},\ \bibinfo {pages} {1} (\bibinfo {year} {1996})}\BibitemShut
  {NoStop}%
\bibitem [{\citenamefont {Leriche}\ \emph {et~al.}(2021)\citenamefont
  {Leriche}, \citenamefont {Palacio-Morales}, \citenamefont {Campetella},
  \citenamefont {Tresca}, \citenamefont {Sasaki}, \citenamefont {Brun},
  \citenamefont {Debontridder}, \citenamefont {David}, \citenamefont {Arfaoui},
  \citenamefont {{\v S}ofranko}, \citenamefont {Samuely}, \citenamefont
  {Kremer}, \citenamefont {Monney}, \citenamefont {Jaouen}, \citenamefont
  {Cario}, \citenamefont {Calandra},\ and\ \citenamefont {Cren}}]{Leriche2021}%
  \BibitemOpen
  \bibfield  {author} {\bibinfo {author} {\bibfnamefont {R.~T.}\ \bibnamefont
  {Leriche}}, \bibinfo {author} {\bibfnamefont {A.}~\bibnamefont
  {Palacio-Morales}}, \bibinfo {author} {\bibfnamefont {M.}~\bibnamefont
  {Campetella}}, \bibinfo {author} {\bibfnamefont {C.}~\bibnamefont {Tresca}},
  \bibinfo {author} {\bibfnamefont {S.}~\bibnamefont {Sasaki}}, \bibinfo
  {author} {\bibfnamefont {C.}~\bibnamefont {Brun}}, \bibinfo {author}
  {\bibfnamefont {F.}~\bibnamefont {Debontridder}}, \bibinfo {author}
  {\bibfnamefont {P.}~\bibnamefont {David}}, \bibinfo {author} {\bibfnamefont
  {I.}~\bibnamefont {Arfaoui}}, \bibinfo {author} {\bibfnamefont
  {O.}~\bibnamefont {{\v S}ofranko}}, \bibinfo {author} {\bibfnamefont
  {T.}~\bibnamefont {Samuely}}, \bibinfo {author} {\bibfnamefont
  {G.}~\bibnamefont {Kremer}}, \bibinfo {author} {\bibfnamefont
  {C.}~\bibnamefont {Monney}}, \bibinfo {author} {\bibfnamefont
  {T.}~\bibnamefont {Jaouen}}, \bibinfo {author} {\bibfnamefont
  {L.}~\bibnamefont {Cario}}, \bibinfo {author} {\bibfnamefont
  {M.}~\bibnamefont {Calandra}},\ and\ \bibinfo {author} {\bibfnamefont
  {T.}~\bibnamefont {Cren}},\ }\bibfield  {title} {\bibinfo {title} {Misfit
  {{Layer Compounds}}: {{A Platform}} for {{Heavily Doped 2D Transition Metal
  Dichalcogenides}}},\ }\href {https://doi.org/10.1002/adfm.202007706}
  {\bibfield  {journal} {\bibinfo  {journal} {Advanced Functional Materials}\
  }\textbf {\bibinfo {volume} {31}},\ \bibinfo {pages} {2007706} (\bibinfo
  {year} {2021})}\BibitemShut {NoStop}%
\bibitem [{\citenamefont {Zullo}\ \emph {et~al.}(2023)\citenamefont {Zullo},
  \citenamefont {Marini}, \citenamefont {Cren},\ and\ \citenamefont
  {Calandra}}]{Zullo2023}%
  \BibitemOpen
  \bibfield  {author} {\bibinfo {author} {\bibfnamefont {L.}~\bibnamefont
  {Zullo}}, \bibinfo {author} {\bibfnamefont {G.}~\bibnamefont {Marini}},
  \bibinfo {author} {\bibfnamefont {T.}~\bibnamefont {Cren}},\ and\ \bibinfo
  {author} {\bibfnamefont {M.}~\bibnamefont {Calandra}},\ }\bibfield  {title}
  {\bibinfo {title} {Misfit {{Layer Compounds}} as {{Ultratunable Field Effect
  Transistors}}: {{From Charge Transfer Control}} to {{Emergent
  Superconductivity}}},\ }\href {https://doi.org/10.1021/acs.nanolett.3c01860}
  {\bibfield  {journal} {\bibinfo  {journal} {Nano Letters}\ }\textbf {\bibinfo
  {volume} {23}},\ \bibinfo {pages} {6658} (\bibinfo {year}
  {2023})}\BibitemShut {NoStop}%
\bibitem [{\citenamefont {Roesky}\ \emph {et~al.}(1993)\citenamefont {Roesky},
  \citenamefont {Meerschaut}, \citenamefont {Rouxel},\ and\ \citenamefont
  {Chen}}]{Roesky1993}%
  \BibitemOpen
  \bibfield  {author} {\bibinfo {author} {\bibfnamefont {R.}~\bibnamefont
  {Roesky}}, \bibinfo {author} {\bibfnamefont {A.}~\bibnamefont {Meerschaut}},
  \bibinfo {author} {\bibfnamefont {J.}~\bibnamefont {Rouxel}},\ and\ \bibinfo
  {author} {\bibfnamefont {J.}~\bibnamefont {Chen}},\ }\bibfield  {title}
  {\bibinfo {title} {Structure and electronic transport properties of the
  {{Misfit}} layer compound
  ({{LaSe}}){\textsubscript{1.14}}({{NbSe}}{\textsubscript{2}}){\textsubscript{2}},
  {{LaNb}}{\textsubscript{2}}{{Se}}{\textsubscript{5}}},\ }\href
  {https://doi.org/10.1002/zaac.19936190119} {\bibfield  {journal} {\bibinfo
  {journal} {Zeitschrift f{\"u}r anorganische und allgemeine Chemie}\ }\textbf
  {\bibinfo {volume} {619}},\ \bibinfo {pages} {117} (\bibinfo {year}
  {1993})}\BibitemShut {NoStop}%
\bibitem [{\citenamefont {Samuely}\ \emph {et~al.}(2021)\citenamefont
  {Samuely}, \citenamefont {Szab{\'o}}, \citenamefont {Ka{\v c}mar{\v
  c}{\'i}k}, \citenamefont {Meerschaut}, \citenamefont {Cario}, \citenamefont
  {Jansen}, \citenamefont {Cren}, \citenamefont {Kuzmiak}, \citenamefont {{\v
  S}ofranko},\ and\ \citenamefont {Samuely}}]{Samuely2021}%
  \BibitemOpen
  \bibfield  {author} {\bibinfo {author} {\bibfnamefont {P.}~\bibnamefont
  {Samuely}}, \bibinfo {author} {\bibfnamefont {P.}~\bibnamefont {Szab{\'o}}},
  \bibinfo {author} {\bibfnamefont {J.}~\bibnamefont {Ka{\v c}mar{\v
  c}{\'i}k}}, \bibinfo {author} {\bibfnamefont {A.}~\bibnamefont {Meerschaut}},
  \bibinfo {author} {\bibfnamefont {L.}~\bibnamefont {Cario}}, \bibinfo
  {author} {\bibfnamefont {A.~G.~M.}\ \bibnamefont {Jansen}}, \bibinfo {author}
  {\bibfnamefont {T.}~\bibnamefont {Cren}}, \bibinfo {author} {\bibfnamefont
  {M.}~\bibnamefont {Kuzmiak}}, \bibinfo {author} {\bibfnamefont
  {O.}~\bibnamefont {{\v S}ofranko}},\ and\ \bibinfo {author} {\bibfnamefont
  {T.}~\bibnamefont {Samuely}},\ }\bibfield  {title} {\bibinfo {title} {Extreme
  in-plane upper critical magnetic fields of heavily doped
  quasi-two-dimensional transition metal dichalcogenides},\ }\href
  {https://doi.org/10.1103/PhysRevB.104.224507} {\bibfield  {journal} {\bibinfo
   {journal} {Physical Review B}\ }\textbf {\bibinfo {volume} {104}},\ \bibinfo
  {pages} {224507} (\bibinfo {year} {2021})}\BibitemShut {NoStop}%
\bibitem [{\citenamefont {Giannozzi}\ \emph {et~al.}(2020)\citenamefont
  {Giannozzi}, \citenamefont {Baseggio}, \citenamefont {Bonf{\`a}},
  \citenamefont {Brunato}, \citenamefont {Car}, \citenamefont {Carnimeo},
  \citenamefont {Cavazzoni}, \citenamefont {De~Gironcoli}, \citenamefont
  {Delugas}, \citenamefont {Ferrari~Ruffino}, \citenamefont {Ferretti},
  \citenamefont {Marzari}, \citenamefont {Timrov}, \citenamefont {Urru},\ and\
  \citenamefont {Baroni}}]{Giannozzi2020}%
  \BibitemOpen
  \bibfield  {author} {\bibinfo {author} {\bibfnamefont {P.}~\bibnamefont
  {Giannozzi}}, \bibinfo {author} {\bibfnamefont {O.}~\bibnamefont {Baseggio}},
  \bibinfo {author} {\bibfnamefont {P.}~\bibnamefont {Bonf{\`a}}}, \bibinfo
  {author} {\bibfnamefont {D.}~\bibnamefont {Brunato}}, \bibinfo {author}
  {\bibfnamefont {R.}~\bibnamefont {Car}}, \bibinfo {author} {\bibfnamefont
  {I.}~\bibnamefont {Carnimeo}}, \bibinfo {author} {\bibfnamefont
  {C.}~\bibnamefont {Cavazzoni}}, \bibinfo {author} {\bibfnamefont
  {S.}~\bibnamefont {De~Gironcoli}}, \bibinfo {author} {\bibfnamefont
  {P.}~\bibnamefont {Delugas}}, \bibinfo {author} {\bibfnamefont
  {F.}~\bibnamefont {Ferrari~Ruffino}}, \bibinfo {author} {\bibfnamefont
  {A.}~\bibnamefont {Ferretti}}, \bibinfo {author} {\bibfnamefont
  {N.}~\bibnamefont {Marzari}}, \bibinfo {author} {\bibfnamefont
  {I.}~\bibnamefont {Timrov}}, \bibinfo {author} {\bibfnamefont
  {A.}~\bibnamefont {Urru}},\ and\ \bibinfo {author} {\bibfnamefont
  {S.}~\bibnamefont {Baroni}},\ }\bibfield  {title} {\bibinfo {title}
  {Q{\textsc{uantum}} {{ESPRESSO}} toward the exascale},\ }\href
  {https://doi.org/10.1063/5.0005082} {\bibfield  {journal} {\bibinfo
  {journal} {The Journal of Chemical Physics}\ }\textbf {\bibinfo {volume}
  {152}},\ \bibinfo {pages} {154105} (\bibinfo {year} {2020})}\BibitemShut
  {NoStop}%
\bibitem [{\citenamefont {Baroni}\ \emph {et~al.}(2001)\citenamefont {Baroni},
  \citenamefont {{de Gironcoli}}, \citenamefont {Dal~Corso},\ and\
  \citenamefont {Giannozzi}}]{Baroni2001}%
  \BibitemOpen
  \bibfield  {author} {\bibinfo {author} {\bibfnamefont {S.}~\bibnamefont
  {Baroni}}, \bibinfo {author} {\bibfnamefont {S.}~\bibnamefont {{de
  Gironcoli}}}, \bibinfo {author} {\bibfnamefont {A.}~\bibnamefont
  {Dal~Corso}},\ and\ \bibinfo {author} {\bibfnamefont {P.}~\bibnamefont
  {Giannozzi}},\ }\bibfield  {title} {\bibinfo {title} {Phonons and related
  crystal properties from density-functional perturbation theory},\ }\href
  {https://doi.org/10.1103/RevModPhys.73.515} {\bibfield  {journal} {\bibinfo
  {journal} {Reviews of Modern Physics}\ }\textbf {\bibinfo {volume} {73}},\
  \bibinfo {pages} {515} (\bibinfo {year} {2001})}\BibitemShut {NoStop}%
\bibitem [{\citenamefont {Perdew}\ \emph {et~al.}(1996)\citenamefont {Perdew},
  \citenamefont {Burke},\ and\ \citenamefont {Ernzerhof}}]{Perdew1996}%
  \BibitemOpen
  \bibfield  {author} {\bibinfo {author} {\bibfnamefont {J.~P.}\ \bibnamefont
  {Perdew}}, \bibinfo {author} {\bibfnamefont {K.}~\bibnamefont {Burke}},\ and\
  \bibinfo {author} {\bibfnamefont {M.}~\bibnamefont {Ernzerhof}},\ }\bibfield
  {title} {\bibinfo {title} {Generalized {{Gradient Approximation Made
  Simple}}},\ }\href {https://doi.org/10.1103/PhysRevLett.77.3865} {\bibfield
  {journal} {\bibinfo  {journal} {Physical Review Letters}\ }\textbf {\bibinfo
  {volume} {77}},\ \bibinfo {pages} {3865} (\bibinfo {year}
  {1996})}\BibitemShut {NoStop}%
\bibitem [{\citenamefont {Grimme}\ \emph {et~al.}(2011)\citenamefont {Grimme},
  \citenamefont {Ehrlich},\ and\ \citenamefont {Goerigk}}]{Grimme2011}%
  \BibitemOpen
  \bibfield  {author} {\bibinfo {author} {\bibfnamefont {S.}~\bibnamefont
  {Grimme}}, \bibinfo {author} {\bibfnamefont {S.}~\bibnamefont {Ehrlich}},\
  and\ \bibinfo {author} {\bibfnamefont {L.}~\bibnamefont {Goerigk}},\
  }\bibfield  {title} {\bibinfo {title} {Effect of the damping function in
  dispersion corrected density functional theory},\ }\href
  {https://doi.org/10.1002/jcc.21759} {\bibfield  {journal} {\bibinfo
  {journal} {Journal of Computational Chemistry}\ }\textbf {\bibinfo {volume}
  {32}},\ \bibinfo {pages} {1456} (\bibinfo {year} {2011})}\BibitemShut
  {NoStop}%
\bibitem [{\citenamefont {Sohier}\ \emph {et~al.}(2017)\citenamefont {Sohier},
  \citenamefont {Calandra},\ and\ \citenamefont {Mauri}}]{Sohier2017}%
  \BibitemOpen
  \bibfield  {author} {\bibinfo {author} {\bibfnamefont {T.}~\bibnamefont
  {Sohier}}, \bibinfo {author} {\bibfnamefont {M.}~\bibnamefont {Calandra}},\
  and\ \bibinfo {author} {\bibfnamefont {F.}~\bibnamefont {Mauri}},\ }\bibfield
   {title} {\bibinfo {title} {Density functional perturbation theory for gated
  two-dimensional heterostructures: {{Theoretical}} developments and
  application to flexural phonons in graphene},\ }\href
  {https://doi.org/10.1103/PhysRevB.96.075448} {\bibfield  {journal} {\bibinfo
  {journal} {Physical Review B}\ }\textbf {\bibinfo {volume} {96}},\ \bibinfo
  {pages} {075448} (\bibinfo {year} {2017})}\BibitemShut {NoStop}%
\bibitem [{\citenamefont {Brumme}\ \emph {et~al.}(2015)\citenamefont {Brumme},
  \citenamefont {Calandra},\ and\ \citenamefont {Mauri}}]{Brumme2015}%
  \BibitemOpen
  \bibfield  {author} {\bibinfo {author} {\bibfnamefont {T.}~\bibnamefont
  {Brumme}}, \bibinfo {author} {\bibfnamefont {M.}~\bibnamefont {Calandra}},\
  and\ \bibinfo {author} {\bibfnamefont {F.}~\bibnamefont {Mauri}},\ }\bibfield
   {title} {\bibinfo {title} {First-principles theory of field-effect doping in
  transition-metal dichalcogenides: {{Structural}} properties, electronic
  structure, {{Hall}} coefficient, and electrical conductivity},\ }\href
  {https://doi.org/10.1103/PhysRevB.91.155436} {\bibfield  {journal} {\bibinfo
  {journal} {Physical Review B}\ }\textbf {\bibinfo {volume} {91}},\ \bibinfo
  {pages} {155436} (\bibinfo {year} {2015})}\BibitemShut {NoStop}%
\bibitem [{\citenamefont {Vanderbilt}(1990)}]{Vanderbilt1990}%
  \BibitemOpen
  \bibfield  {author} {\bibinfo {author} {\bibfnamefont {D.}~\bibnamefont
  {Vanderbilt}},\ }\bibfield  {title} {\bibinfo {title} {Soft self-consistent
  pseudopotentials in a generalized eigenvalue formalism},\ }\href
  {https://doi.org/10.1103/PhysRevB.41.7892} {\bibfield  {journal} {\bibinfo
  {journal} {Physical Review B}\ }\textbf {\bibinfo {volume} {41}},\ \bibinfo
  {pages} {7892} (\bibinfo {year} {1990})}\BibitemShut {NoStop}%
\bibitem [{\citenamefont {Bianco}\ \emph {et~al.}(2020)\citenamefont {Bianco},
  \citenamefont {Monacelli}, \citenamefont {Calandra}, \citenamefont {Mauri},\
  and\ \citenamefont {Errea}}]{Bianco2020}%
  \BibitemOpen
  \bibfield  {author} {\bibinfo {author} {\bibfnamefont {R.}~\bibnamefont
  {Bianco}}, \bibinfo {author} {\bibfnamefont {L.}~\bibnamefont {Monacelli}},
  \bibinfo {author} {\bibfnamefont {M.}~\bibnamefont {Calandra}}, \bibinfo
  {author} {\bibfnamefont {F.}~\bibnamefont {Mauri}},\ and\ \bibinfo {author}
  {\bibfnamefont {I.}~\bibnamefont {Errea}},\ }\bibfield  {title} {\bibinfo
  {title} {Weak {{Dimensionality Dependence}} and {{Dominant Role}} of {{Ionic
  Fluctuations}} in the {{Charge-Density-Wave Transition}} of {{NbSe}}$_{2}$},\
  }\href {https://doi.org/10.1103/PhysRevLett.125.106101} {\bibfield  {journal}
  {\bibinfo  {journal} {Physical Review Letters}\ }\textbf {\bibinfo {volume}
  {125}},\ \bibinfo {pages} {106101} (\bibinfo {year} {2020})}\BibitemShut
  {NoStop}%
\bibitem [{\citenamefont {Sugai}\ \emph {et~al.}(1981)\citenamefont {Sugai},
  \citenamefont {Murase}, \citenamefont {Uchida},\ and\ \citenamefont
  {Tanaka}}]{Sugai1981}%
  \BibitemOpen
  \bibfield  {author} {\bibinfo {author} {\bibfnamefont {S.}~\bibnamefont
  {Sugai}}, \bibinfo {author} {\bibfnamefont {K.}~\bibnamefont {Murase}},
  \bibinfo {author} {\bibfnamefont {S.}~\bibnamefont {Uchida}},\ and\ \bibinfo
  {author} {\bibfnamefont {S.}~\bibnamefont {Tanaka}},\ }\bibfield  {title}
  {\bibinfo {title} {Comparison of the soft modes in tantalum
  dichalcogenides},\ }\href {https://doi.org/10.1016/0378-4363(81)90284-9}
  {\bibfield  {journal} {\bibinfo  {journal} {Physica B+C}\ }\textbf {\bibinfo
  {volume} {105}},\ \bibinfo {pages} {405} (\bibinfo {year}
  {1981})}\BibitemShut {NoStop}%
\bibitem [{\citenamefont {Tsang}\ \emph {et~al.}(1976)\citenamefont {Tsang},
  \citenamefont {Smith},\ and\ \citenamefont {Shafer}}]{Tsang1976}%
  \BibitemOpen
  \bibfield  {author} {\bibinfo {author} {\bibfnamefont {J.~C.}\ \bibnamefont
  {Tsang}}, \bibinfo {author} {\bibfnamefont {J.~E.}\ \bibnamefont {Smith}},\
  and\ \bibinfo {author} {\bibfnamefont {M.~W.}\ \bibnamefont {Shafer}},\
  }\bibfield  {title} {\bibinfo {title} {Raman {{Spectroscopy}} of {{Soft
  Modes}} at the {{Charge-Density-Wave Phase Transition}} in 2 {{H}} -
  {{NbSe}}$_{2}$},\ }\href {https://doi.org/10.1103/PhysRevLett.37.1407}
  {\bibfield  {journal} {\bibinfo  {journal} {Physical Review Letters}\
  }\textbf {\bibinfo {volume} {37}},\ \bibinfo {pages} {1407} (\bibinfo {year}
  {1976})}\BibitemShut {NoStop}%
\bibitem [{\citenamefont {M{\'e}asson}\ \emph {et~al.}(2014)\citenamefont
  {M{\'e}asson}, \citenamefont {Gallais}, \citenamefont {Cazayous},
  \citenamefont {Clair}, \citenamefont {Rodi{\`e}re}, \citenamefont {Cario},\
  and\ \citenamefont {Sacuto}}]{Measson2014}%
  \BibitemOpen
  \bibfield  {author} {\bibinfo {author} {\bibfnamefont {M.-A.}\ \bibnamefont
  {M{\'e}asson}}, \bibinfo {author} {\bibfnamefont {Y.}~\bibnamefont
  {Gallais}}, \bibinfo {author} {\bibfnamefont {M.}~\bibnamefont {Cazayous}},
  \bibinfo {author} {\bibfnamefont {B.}~\bibnamefont {Clair}}, \bibinfo
  {author} {\bibfnamefont {P.}~\bibnamefont {Rodi{\`e}re}}, \bibinfo {author}
  {\bibfnamefont {L.}~\bibnamefont {Cario}},\ and\ \bibinfo {author}
  {\bibfnamefont {A.}~\bibnamefont {Sacuto}},\ }\bibfield  {title} {\bibinfo
  {title} {Amplitude {{Higgs}} mode in the {{2H}}-{{NbSe$_{2}$}}
  superconductor},\ }\href {https://doi.org/10.1103/PhysRevB.89.060503}
  {\bibfield  {journal} {\bibinfo  {journal} {Physical Review B}\ }\textbf
  {\bibinfo {volume} {89}},\ \bibinfo {pages} {060503} (\bibinfo {year}
  {2014})}\BibitemShut {NoStop}%
\bibitem [{\citenamefont {Grasset}\ \emph {et~al.}(2018)\citenamefont
  {Grasset}, \citenamefont {Cea}, \citenamefont {Gallais}, \citenamefont
  {Cazayous}, \citenamefont {Sacuto}, \citenamefont {Cario}, \citenamefont
  {Benfatto},\ and\ \citenamefont {M{\'e}asson}}]{Grasset2018a}%
  \BibitemOpen
  \bibfield  {author} {\bibinfo {author} {\bibfnamefont {R.}~\bibnamefont
  {Grasset}}, \bibinfo {author} {\bibfnamefont {T.}~\bibnamefont {Cea}},
  \bibinfo {author} {\bibfnamefont {Y.}~\bibnamefont {Gallais}}, \bibinfo
  {author} {\bibfnamefont {M.}~\bibnamefont {Cazayous}}, \bibinfo {author}
  {\bibfnamefont {A.}~\bibnamefont {Sacuto}}, \bibinfo {author} {\bibfnamefont
  {L.}~\bibnamefont {Cario}}, \bibinfo {author} {\bibfnamefont
  {L.}~\bibnamefont {Benfatto}},\ and\ \bibinfo {author} {\bibfnamefont
  {M.-A.}\ \bibnamefont {M{\'e}asson}},\ }\bibfield  {title} {\bibinfo {title}
  {Higgs-mode radiance and charge-density-wave order in
  {{2H}}-{{NbSe$_{2}$}}},\ }\href {https://doi.org/10.1103/PhysRevB.97.094502}
  {\bibfield  {journal} {\bibinfo  {journal} {Physical Review B}\ }\textbf
  {\bibinfo {volume} {97}},\ \bibinfo {pages} {094502} (\bibinfo {year}
  {2018})}\BibitemShut {NoStop}%
\bibitem [{\citenamefont {Grasset}\ \emph {et~al.}(2019)\citenamefont
  {Grasset}, \citenamefont {Gallais}, \citenamefont {Sacuto}, \citenamefont
  {Cazayous}, \citenamefont {{Ma{\~n}as-Valero}}, \citenamefont {Coronado},\
  and\ \citenamefont {M{\'e}asson}}]{Grasset2019}%
  \BibitemOpen
  \bibfield  {author} {\bibinfo {author} {\bibfnamefont {R.}~\bibnamefont
  {Grasset}}, \bibinfo {author} {\bibfnamefont {Y.}~\bibnamefont {Gallais}},
  \bibinfo {author} {\bibfnamefont {A.}~\bibnamefont {Sacuto}}, \bibinfo
  {author} {\bibfnamefont {M.}~\bibnamefont {Cazayous}}, \bibinfo {author}
  {\bibfnamefont {S.}~\bibnamefont {{Ma{\~n}as-Valero}}}, \bibinfo {author}
  {\bibfnamefont {E.}~\bibnamefont {Coronado}},\ and\ \bibinfo {author}
  {\bibfnamefont {M.-A.}\ \bibnamefont {M{\'e}asson}},\ }\bibfield  {title}
  {\bibinfo {title} {Pressure-{{Induced Collapse}} of the {{Charge Density
  Wave}} and {{Higgs Mode Visibility}} in 2 {{H}}-{{TaS}}$_{2}$},\ }\href
  {https://doi.org/10.1103/PhysRevLett.122.127001} {\bibfield  {journal}
  {\bibinfo  {journal} {Physical Review Letters}\ }\textbf {\bibinfo {volume}
  {122}},\ \bibinfo {pages} {127001} (\bibinfo {year} {2019})}\BibitemShut
  {NoStop}%
\bibitem [{\citenamefont {He}\ \emph {et~al.}(2024)\citenamefont {He},
  \citenamefont {Peis}, \citenamefont {Cuddy}, \citenamefont {Zhao},
  \citenamefont {Li}, \citenamefont {Zhang}, \citenamefont {Stumberger},
  \citenamefont {Moritz}, \citenamefont {Yang}, \citenamefont {Gao},
  \citenamefont {Devereaux},\ and\ \citenamefont {Hackl}}]{He2024}%
  \BibitemOpen
  \bibfield  {author} {\bibinfo {author} {\bibfnamefont {G.}~\bibnamefont
  {He}}, \bibinfo {author} {\bibfnamefont {L.}~\bibnamefont {Peis}}, \bibinfo
  {author} {\bibfnamefont {E.~F.}\ \bibnamefont {Cuddy}}, \bibinfo {author}
  {\bibfnamefont {Z.}~\bibnamefont {Zhao}}, \bibinfo {author} {\bibfnamefont
  {D.}~\bibnamefont {Li}}, \bibinfo {author} {\bibfnamefont {Y.}~\bibnamefont
  {Zhang}}, \bibinfo {author} {\bibfnamefont {R.}~\bibnamefont {Stumberger}},
  \bibinfo {author} {\bibfnamefont {B.}~\bibnamefont {Moritz}}, \bibinfo
  {author} {\bibfnamefont {H.}~\bibnamefont {Yang}}, \bibinfo {author}
  {\bibfnamefont {H.}~\bibnamefont {Gao}}, \bibinfo {author} {\bibfnamefont
  {T.~P.}\ \bibnamefont {Devereaux}},\ and\ \bibinfo {author} {\bibfnamefont
  {R.}~\bibnamefont {Hackl}},\ }\bibfield  {title} {\bibinfo {title}
  {Anharmonic strong-coupling effects at the origin of the charge density wave
  in {{CsV$_{3}$Sb$_{5}$}}},\ }\href
  {https://doi.org/10.1038/s41467-024-45865-0} {\bibfield  {journal} {\bibinfo
  {journal} {Nature Communications}\ }\textbf {\bibinfo {volume} {15}},\
  \bibinfo {pages} {1895} (\bibinfo {year} {2024})}\BibitemShut {NoStop}%
\bibitem [{\citenamefont {Lin}\ \emph {et~al.}(2020)\citenamefont {Lin},
  \citenamefont {Li}, \citenamefont {Wen}, \citenamefont {Berger},
  \citenamefont {Forr{\'o}}, \citenamefont {Zhou}, \citenamefont {Jia},
  \citenamefont {Taniguchi}, \citenamefont {Watanabe}, \citenamefont {Xi},\
  and\ \citenamefont {Bahramy}}]{Lin2020}%
  \BibitemOpen
  \bibfield  {author} {\bibinfo {author} {\bibfnamefont {D.}~\bibnamefont
  {Lin}}, \bibinfo {author} {\bibfnamefont {S.}~\bibnamefont {Li}}, \bibinfo
  {author} {\bibfnamefont {J.}~\bibnamefont {Wen}}, \bibinfo {author}
  {\bibfnamefont {H.}~\bibnamefont {Berger}}, \bibinfo {author} {\bibfnamefont
  {L.}~\bibnamefont {Forr{\'o}}}, \bibinfo {author} {\bibfnamefont
  {H.}~\bibnamefont {Zhou}}, \bibinfo {author} {\bibfnamefont {S.}~\bibnamefont
  {Jia}}, \bibinfo {author} {\bibfnamefont {T.}~\bibnamefont {Taniguchi}},
  \bibinfo {author} {\bibfnamefont {K.}~\bibnamefont {Watanabe}}, \bibinfo
  {author} {\bibfnamefont {X.}~\bibnamefont {Xi}},\ and\ \bibinfo {author}
  {\bibfnamefont {M.~S.}\ \bibnamefont {Bahramy}},\ }\bibfield  {title}
  {\bibinfo {title} {Patterns and driving forces of dimensionality-dependent
  charge density waves in {{2H-type}} transition metal dichalcogenides},\
  }\href {https://doi.org/10.1038/s41467-020-15715-w} {\bibfield  {journal}
  {\bibinfo  {journal} {Nature Communications}\ }\textbf {\bibinfo {volume}
  {11}},\ \bibinfo {pages} {2406} (\bibinfo {year} {2020})}\BibitemShut
  {NoStop}%
\bibitem [{\citenamefont {Sooryakumar}\ \emph {et~al.}(1981)\citenamefont
  {Sooryakumar}, \citenamefont {Klein},\ and\ \citenamefont
  {Frindt}}]{Sooryakumar1981a}%
  \BibitemOpen
  \bibfield  {author} {\bibinfo {author} {\bibfnamefont {R.}~\bibnamefont
  {Sooryakumar}}, \bibinfo {author} {\bibfnamefont {M.~V.}\ \bibnamefont
  {Klein}},\ and\ \bibinfo {author} {\bibfnamefont {R.~F.}\ \bibnamefont
  {Frindt}},\ }\bibfield  {title} {\bibinfo {title} {Effect of nonmagnetic
  impurities on the {{Raman}} spectra of the superconductor niobium
  diselenide},\ }\href {https://doi.org/10.1103/PhysRevB.23.3222} {\bibfield
  {journal} {\bibinfo  {journal} {Physical Review B}\ }\textbf {\bibinfo
  {volume} {23}},\ \bibinfo {pages} {3222} (\bibinfo {year}
  {1981})}\BibitemShut {NoStop}%
\bibitem [{\citenamefont {Arguello}\ \emph {et~al.}(2014)\citenamefont
  {Arguello}, \citenamefont {Chockalingam}, \citenamefont {Rosenthal},
  \citenamefont {Zhao}, \citenamefont {Guti{\'e}rrez}, \citenamefont {Kang},
  \citenamefont {Chung}, \citenamefont {Fernandes}, \citenamefont {Jia},
  \citenamefont {Millis}, \citenamefont {Cava},\ and\ \citenamefont
  {Pasupathy}}]{Arguello2014}%
  \BibitemOpen
  \bibfield  {author} {\bibinfo {author} {\bibfnamefont {C.~J.}\ \bibnamefont
  {Arguello}}, \bibinfo {author} {\bibfnamefont {S.~P.}\ \bibnamefont
  {Chockalingam}}, \bibinfo {author} {\bibfnamefont {E.~P.}\ \bibnamefont
  {Rosenthal}}, \bibinfo {author} {\bibfnamefont {L.}~\bibnamefont {Zhao}},
  \bibinfo {author} {\bibfnamefont {C.}~\bibnamefont {Guti{\'e}rrez}}, \bibinfo
  {author} {\bibfnamefont {J.~H.}\ \bibnamefont {Kang}}, \bibinfo {author}
  {\bibfnamefont {W.~C.}\ \bibnamefont {Chung}}, \bibinfo {author}
  {\bibfnamefont {R.~M.}\ \bibnamefont {Fernandes}}, \bibinfo {author}
  {\bibfnamefont {S.}~\bibnamefont {Jia}}, \bibinfo {author} {\bibfnamefont
  {A.~J.}\ \bibnamefont {Millis}}, \bibinfo {author} {\bibfnamefont {R.~J.}\
  \bibnamefont {Cava}},\ and\ \bibinfo {author} {\bibfnamefont {A.~N.}\
  \bibnamefont {Pasupathy}},\ }\bibfield  {title} {\bibinfo {title}
  {Visualizing the charge density wave transition in {{2H}}-{{NbSe}}$_{2}$ in
  real space},\ }\bibfield  {journal} {\bibinfo  {journal} {Physical Review B}\
  }\textbf {\bibinfo {volume} {89}},\ \href
  {https://doi.org/10.1103/PhysRevB.89.235115} {10.1103/PhysRevB.89.235115}
  (\bibinfo {year} {2014})\BibitemShut {NoStop}%
\bibitem [{\citenamefont {Oh}\ \emph {et~al.}(2020)\citenamefont {Oh},
  \citenamefont {Gye},\ and\ \citenamefont {Yeom}}]{Oh2020}%
  \BibitemOpen
  \bibfield  {author} {\bibinfo {author} {\bibfnamefont {E.}~\bibnamefont
  {Oh}}, \bibinfo {author} {\bibfnamefont {G.}~\bibnamefont {Gye}},\ and\
  \bibinfo {author} {\bibfnamefont {H.~W.}\ \bibnamefont {Yeom}},\ }\bibfield
  {title} {\bibinfo {title} {Defect-{{Selective Charge-Density-Wave
  Condensation}} in 2{{H}}-{{NbSe}}$_{2}$},\ }\href
  {https://doi.org/10.1103/PhysRevLett.125.036804} {\bibfield  {journal}
  {\bibinfo  {journal} {Physical Review Letters}\ }\textbf {\bibinfo {volume}
  {125}},\ \bibinfo {pages} {036804} (\bibinfo {year} {2020})}\BibitemShut
  {NoStop}%
\bibitem [{\citenamefont {Sahoo}\ \emph {et~al.}(2022)\citenamefont {Sahoo},
  \citenamefont {Mukherjee},\ and\ \citenamefont {Sahoo}}]{Sahoo2022}%
  \BibitemOpen
  \bibfield  {author} {\bibinfo {author} {\bibfnamefont {U.~P.}\ \bibnamefont
  {Sahoo}}, \bibinfo {author} {\bibfnamefont {A.}~\bibnamefont {Mukherjee}},\
  and\ \bibinfo {author} {\bibfnamefont {P.~K.}\ \bibnamefont {Sahoo}},\
  }\bibfield  {title} {\bibinfo {title} {Short-{{Range Charge Density Wave}}
  and {{Bandgap Modulation}} by {{Au-Implanted Defects}} in
  {{TiSe}}{\textsubscript{2}}},\ }\href
  {https://doi.org/10.1021/acsaelm.2c00287} {\bibfield  {journal} {\bibinfo
  {journal} {ACS Applied Electronic Materials}\ }\textbf {\bibinfo {volume}
  {4}},\ \bibinfo {pages} {3428} (\bibinfo {year} {2022})}\BibitemShut
  {NoStop}%
\bibitem [{\citenamefont {Brun}\ \emph {et~al.}(2010)\citenamefont {Brun},
  \citenamefont {Wang}, \citenamefont {Monceau},\ and\ \citenamefont
  {Brazovskii}}]{Brun2010}%
  \BibitemOpen
  \bibfield  {author} {\bibinfo {author} {\bibfnamefont {C.}~\bibnamefont
  {Brun}}, \bibinfo {author} {\bibfnamefont {Z.-Z.}\ \bibnamefont {Wang}},
  \bibinfo {author} {\bibfnamefont {P.}~\bibnamefont {Monceau}},\ and\ \bibinfo
  {author} {\bibfnamefont {S.}~\bibnamefont {Brazovskii}},\ }\bibfield  {title}
  {\bibinfo {title} {Surface {{Charge Density Wave Phase Transition}} in
  {{NbS}}$_{3}$},\ }\href {https://doi.org/10.1103/PhysRevLett.104.256403}
  {\bibfield  {journal} {\bibinfo  {journal} {Physical Review Letters}\
  }\textbf {\bibinfo {volume} {104}},\ \bibinfo {pages} {256403} (\bibinfo
  {year} {2010})}\BibitemShut {NoStop}%
\bibitem [{\citenamefont {{Machado-Charry}}\ \emph {et~al.}(2006)\citenamefont
  {{Machado-Charry}}, \citenamefont {Ordej{\'o}n}, \citenamefont {Canadell},
  \citenamefont {Brun},\ and\ \citenamefont {Wang}}]{Machado-Charry2006}%
  \BibitemOpen
  \bibfield  {author} {\bibinfo {author} {\bibfnamefont {E.}~\bibnamefont
  {{Machado-Charry}}}, \bibinfo {author} {\bibfnamefont {P.}~\bibnamefont
  {Ordej{\'o}n}}, \bibinfo {author} {\bibfnamefont {E.}~\bibnamefont
  {Canadell}}, \bibinfo {author} {\bibfnamefont {C.}~\bibnamefont {Brun}},\
  and\ \bibinfo {author} {\bibfnamefont {Z.~Z.}\ \bibnamefont {Wang}},\
  }\bibfield  {title} {\bibinfo {title} {Analysis of scanning tunneling
  microscopy images of the charge-density-wave phase in quasi-one-dimensional
  {{Rb}}$_{0.3}${{MoO}}$_{3}$},\ }\href
  {https://doi.org/10.1103/PhysRevB.74.155123} {\bibfield  {journal} {\bibinfo
  {journal} {Physical Review B}\ }\textbf {\bibinfo {volume} {74}},\ \bibinfo
  {pages} {155123} (\bibinfo {year} {2006})}\BibitemShut {NoStop}%
\bibitem [{\citenamefont {Brun}\ \emph {et~al.}(2007)\citenamefont {Brun},
  \citenamefont {{Machado-Charry}}, \citenamefont {Ordej{\'o}n}, \citenamefont
  {Canadell},\ and\ \citenamefont {Wang}}]{Brun2007}%
  \BibitemOpen
  \bibfield  {author} {\bibinfo {author} {\bibfnamefont {C.}~\bibnamefont
  {Brun}}, \bibinfo {author} {\bibfnamefont {E.}~\bibnamefont
  {{Machado-Charry}}}, \bibinfo {author} {\bibfnamefont {P.}~\bibnamefont
  {Ordej{\'o}n}}, \bibinfo {author} {\bibfnamefont {E.}~\bibnamefont
  {Canadell}},\ and\ \bibinfo {author} {\bibfnamefont {Z.~Z.}\ \bibnamefont
  {Wang}},\ }\bibfield  {title} {\bibinfo {title} {Inhomogenities of the
  {{CDW}} vector at the (-201) surface of {{Quasi-1D}} blue bronze
  {{Rb}}{\textsubscript{0.3}}{{MoO}}{\textsubscript{3}}},\ }\href
  {https://doi.org/10.1088/1742-6596/61/1/029} {\bibfield  {journal} {\bibinfo
  {journal} {Journal of Physics: Conference Series}\ }\textbf {\bibinfo
  {volume} {61}},\ \bibinfo {pages} {140} (\bibinfo {year} {2007})}\BibitemShut
  {NoStop}%
\end{thebibliography}%



\end{document}